\documentclass[twocolumn]{aastex62}

\usepackage{booktabs}
\usepackage{color}
\usepackage{amsmath}
\usepackage{hyperref}
\usepackage{graphicx}
\usepackage{dcolumn}
\usepackage{bm}
\definecolor{amber}{rgb}{1.0, 0.75, 0.0}

\newcommand{\Fermi}{\textit{Fermi}}

\shorttitle{Galactic Millisecond Pulsars}
\shortauthors{R.~T.~Bartels, T.~D.~P.~Edwards and C.~Weniger}

\begin{document}

\title{Bayesian Model Comparison and Analysis of the Galactic Disk Population of Gamma-Ray Millisecond Pulsars}

\author{R.~T.~Bartels}
\email{r.t.bartels@uva.nl}
\affiliation{Gravitation Astroparticle Physics Amsterdam (GRAPPA), Institute
for Theoretical Physics Amsterdam and Delta Institute for Theoretical Physics,
University of Amsterdam, Science Park 904, 1090 GL Amsterdam, The Netherlands}
\author{T.~D.~P.~Edwards}
\email{t.d.p.edwards@uva.nl}
\affiliation{Gravitation Astroparticle Physics Amsterdam (GRAPPA), Institute
for Theoretical Physics Amsterdam and Delta Institute for Theoretical Physics,
University of Amsterdam, Science Park 904, 1090 GL Amsterdam, The Netherlands}
\author{C. Weniger}
\email{c.weniger@uva.nl}
\affiliation{Gravitation Astroparticle Physics Amsterdam (GRAPPA), Institute
for Theoretical Physics Amsterdam and Delta Institute for Theoretical Physics,
University of Amsterdam, Science Park 904, 1090 GL Amsterdam, The Netherlands}
\date{\today}

\begin{abstract}
Pulsed emission from almost one hundred millisecond pulsars (MSPs) has been detected 
in $\gamma$-rays by the \Fermi\ Large-Area Telescope.
The global properties of this population remain relatively unconstrained despite many attempts to model their spatial and luminosity distributions. We perform here a self-consistent Bayesian analysis of both the spatial distribution and luminosity function simultaneously. Distance uncertainties, arising from errors in the parallax measurement or Galactic electron-density model, are marginalized over.
We provide a public \texttt{Python} 
package\footnote{Available from \url{http://github.com/tedwards2412/MSPDist}.} for calculating distance uncertainties to pulsars derived using the dispersion measure by accounting for the uncertainties in Galactic electron-density model YMW16. 
Finally, we use multiple parameterizations for the MSP population and perform Bayesian model comparison, finding that a broken power law luminosity function with Lorimer spatial profile are preferred over multiple other parameterizations used in the past.
The best-fit spatial distribution and number of $\gamma$-ray MSPs is consistent
with results for the radio population of MSPs.
\end{abstract}

\keywords{pulsars: general --- gamma-rays: general --- stars: luminosity function --- Galaxy: disk}

\section{\label{sec:Intro}Introduction}

Millisecond pulsars (MSPs) are believed to be recycled pulsars that are spun-up to millisecond periods
by accreting matter from a companion star \citep{1991PhR...203....1B}.
Prior to the launch of the \Fermi\ Gamma-Ray Space Telescope pulsations from only one MSP had been claimed in 
$\gamma$-rays and at low statistical significance \citep{Kuiper:2000ce}.
Since then the Large Area Telescope (LAT) aboard \Fermi\ has revolutionized the field
with close to one hundred $\gamma$-ray detected millisecond pulsars
\citep{Caraveo:2013lra, Abdo:2009fn,Abdo:2009jfa,TheFermi-LAT:2013ssa}.
Most detections of $\gamma$-ray pulsations in MSPs follow from phase-folding the
timing parameters already known from radio \citep[e.g.~][]{Abdo:2009jfa}.
In many cases, the radio MSPs have been initially detected during follow-up observations
of \Fermi\ unassociated sources after which the timing information is utilized to confirm 
$\gamma$-ray pulsations \citep[e.g.~][]{Cognard:2011sf}.
Increased computing power has made it possible to detect
$\gamma$-ray pulsations in blind searches where no timing information is available
\citep{Pletsch:2012sh,Clark:2018zsp}.

Population studies of MSPs in radio have constrained their spatial distribution, luminosity function
and the
number of radio-emitting MSPs in the Galactic disk \citep{Cordes:1997my, 1998MNRAS.295..743L, Levin:2013usa}.
On the other hand, $\gamma$-ray population studies of MSPs have been performed to constrain their 
luminosity function and in some cases their spatial distribution
\citep{Hooper:2013nhl, Gregoire:2013yta, Yuan:2014rca, Cholis:2014noa, Hooper:2015jlu, Winter:2016wmy, Ploeg:2017vai}.

A particular goal of many of these analyses has been to rule-out or constrain the MSP
interpretation of the \Fermi\ Galactic Center Excess (GCE).
The GCE is an excess of $\gamma$-rays at energies of $\sim 2\mathrm{\,GeV}$
that is spatially coincident with the Galactic Bulge
\citep{Goodenough:2009gk, Daylan:2014rsa, Calore:2014xka} and
was also shown to be morphologically similar \citep{Bartels:2017vsx}.
It has been suggested that the GCE could be caused by a bulge population of MSPs
\citep{Abazajian:2010zy, Gordon:2013vta}. Corroborative evidence
for this scenario was found by analysing the photon statistics of the inner-Galaxy
\citep{Lee:2015fea, Bartels:2015aea}.
However, arguments against this scenario exist based on an apparent
conflict between the luminosity function of MSPs in the Galactic disk and the intensity of 
the GCE. It was argued that if the GCE is caused by MSPs we should have already
detected a few dozen sources from this population \citep{Hooper:2013nhl,Cholis:2014noa, Hooper:2015jlu}. 
Conversely, other studies claimed that there is no discrepancy if bulge MSPs
have the same luminosity function as disk MSPs \citep{Yuan:2014rca,Petrovic:2014xra,Ploeg:2017vai}.
Previous analyses have used a variety of distributions for the luminosity
function of MSPs. Moreover, they have used different treatments of the distance estimates
to MSPs, which is one of the major sources of uncertainty when estimating the pulsar luminosity. In the light of conflicting conclusions caused by particular assumptions it seems important to perform a complete and unbiased analysis, presenting all sources of uncertainty clearly and adopting a conservative set of assumptions. 

\medskip

In this work we perform a systematic and fully self-consistent analyses of the spatial
distribution and luminosity function of MSPs. We consider different luminosity functions
and parameterizations of the spatial profile, performing a Bayesian-unbinned likelihood analysis to constrain the model parameters.
Bayesian model comparison is then applied to select the best model.
In our analysis we marginalize over the main sources of uncertainty, namely the distance to and received flux of each source. 
What is more, to the best of our knowledge, we for the first time construct probability distribution
functions for distances derived from the dispersion measure by taking into account
the uncertainties in the parameters of the electron-density models \citep{2017ApJ...835...29Y}.
Finally, we also study how the inclusion of unassociated sources can impact our results.

\medskip

The layout of the paper is as follows. We first discuss our modeling and MSP 
data sample in Sec.~\ref{sec:methods}. Results are then given in Sect.~\ref{sec:results}.  Finally we discuss the implications of our results in Sect.~\ref{sec:discussion} and conclude in Sect.~\ref{sec:conclusion}.

\section{\label{sec:methods}Methodology and Data}
In this work we perform a Bayesian-unbinned likelihood analysis in order to fully exploit the heterogeneous information available in the data sample.  We first discuss the likelihood function and then address the two main areas of uncertainty, namely the distances to sources and the contribution from unassociated sources.

\medskip

\subsection{Likelihood}
Our analysis is based on an unbinned Poisson likelihood function,
\begin{equation}
\label{eq:likelihood}
    \mathcal{L}\left(\mathcal{D}|\mathbf{\Theta}\right) = e^{-\mu(\mathbf{\Theta})}\prod_i^{N_\mathrm{obs}}N_\mathrm{tot} P(\mathcal{D}_i|\mathbf{\Theta})\; ,
\end{equation}
where $\mathbf{\Theta}$ is the vector of parameter dependencies,
$N_\mathrm{obs}$ is the number of observed MSPs, $N_\mathrm{tot}$ the total number of sources and $P(\mathcal{D}_i|\mathbf{\Theta})$ is the probability of finding a given source at Galactic position $(\ell_i, b_i)$, with observed flux $F_i$ and, if available, parallax or dispersion measure $\omega_i$ or $\mathrm{DM}_i$, 
i.e.~$\mathcal{D}_i=\left\{\ell_i, b_i, F_i, \kappa_i\right\}$ with
$\kappa_i = \omega_i$ if a parallax measurement is present, or else
$\kappa_i = \mathrm{DM}_i$ if a DM measurement exists. 
If no distance measure is present $\mathcal{D}_i=\left\{\ell_i, b_i, F_i\right\}$. 
Furthermore, $\mu(\mathbf{\Theta})$ is the expected number of observed sources and satisfies in the point of maximum likelihood the condition
\begin{equation}
   \mu(\mathbf{\Theta}_{\mathrm{bf}}) = N_{\text{obs}}\;,
\end{equation}
where $\mathbf{\Theta}_{\mathrm{bf}}$ are the maximum-likelihood values for the parameters of our model.
More specifically, $\mu(\mathbf{\Theta})$ and $P(\mathcal{D}_i|\mathbf{\Theta})$ are given by
\begin{widetext}
\begin{equation}
\label{eq:mu}
\begin{split}
  \mu(\mathbf{\Theta}) = N_\mathrm{tot}
  \sum_{j=1}^\mathrm{N_\mathrm{pix}}\frac{\Omega_j}{\cos{b_j}}
  \int dD\int dL \,P(L|\mathbf{\Theta}) P(\ell_j, b_j, D|\mathbf{\Theta})
  P_\mathrm{th}\left(\frac{L}{4\pi D^2} \Big|
  \mathbf{\Theta}, \ell_j, b_j\right),
\end{split}
\end{equation}
\begin{subequations}
\label{eq:prob}
  \begin{equation}
  \label{eq:prob_dist}
    P(\ell_i, b_i, F_i, \kappa_i|\mathbf{\Theta}) = 4 \pi \int dD \int dF\,
    D^2
    P(\ell_i, b_i, D|\mathbf{\Theta}) 
    P\left(L=4\pi D^2 F|\mathbf{\Theta}\right)
    P_\mathrm{th}\left(F\right|\mathbf{\Theta}, \ell_i, b_i)
    P(\kappa_i | D) P(F_i |F)
  \end{equation}
  \begin{equation}
  \label{eq:prob_nodist}
    P(\ell_i, b_i,F_i|\mathbf{\Theta}) = 4 \pi \int dD \int dF\,
    D^2
    P(\ell_i, b_i, D|\mathbf{\Theta}) 
    P\left(L=4\pi D^2 F|\mathbf{\Theta}\right)
    P_\mathrm{th}\left(F\right|\mathbf{\Theta}, \ell_i, b_i)
    P(F_i |F)
  \end{equation}
\end{subequations}
\end{widetext}
Here $P(L|\mathbf{\Theta})$ and $P(\ell, b, D|\mathbf{\Theta})$ are the luminosity
function and spatial distribution, which are discussed in detail in 
sections \ref{sec:L} and \ref{sec:rho} respectively. 
The total number of sources equals the sum of the disk ($N$)
and bulge sources ($N_\mathrm{bulge}$), $N_\mathrm{tot}=N + N_\mathrm{bulge}$.
$P_\mathrm{th}(F|\mathbf{\Theta}, \ell, b)$
is the detection sensitivity which is defined in Eq.~\ref{eq:Pth}.
We take the observed spatial positions to correspond to the true positions, 
since their uncertainties are negligible for the purpose of our analysis.
On the other hand, we integrate over the true distances ($D$) and fluxes ($F$) of the sources.
In Sect.~\ref{sec:flux} we discuss $P(F_i|F)$, 
the probability of measuring a flux ($F_i$), given the true flux
of the source ($F$).
Similarly,
$P(\kappa_i|D)$ is the probability of observing a particular parallax or dispersion measure value
($\kappa_i$)
given a true distance to the source. It is discussed separately in Sect.~\ref{sec:dist}.
Equation \ref{eq:prob_dist} (\ref{eq:prob_nodist}) applies to sources with (without)
distance information.

In order to compute the expected number of observed sources $\mu$ we must integrate over distance, flux and spatial coordinates.
The spatial integral is performed by calculating expectations on a \texttt{HEALPIX} grid 
with $\mathrm{NSIDE}=32$ \citep{Gorski:2004by}.  In this case the number of pixels is $N_\mathrm{pix}=12288$ and
$\Omega_i = 1\times 10^{-3} \mathrm{\,sr}$.  The integral is then straightforwardly performed
by summing over all pixels. Since we integrate over solid angle rather than
$\ell, b$ we divide out a factor of $\cos{b}$ in
Eq.~\ref{eq:mu} which appears in $P(\ell, b, D)$.
Henceforth, we drop the dependence on $\mathbf{\Theta}$ for notational purposes but note that the free parameters are clearly stated in Table.~\ref{tab:params}.

\subsubsection{Luminosity function}
\label{sec:L}

We test four parameterizations of the luminosity function
in the range $0.1$--$100\mathrm{\,GeV}$, namely a single power law with a hard cutoff 
(PL, Eq.~\ref{eq:PL}), 
single power law with super-exponential cutoff (PL exp. cutoff, Eq.~\ref{eq:PLexp}),
broken power law (BPL, Eq.~\ref{eq:BPL})
and log-normal distribution (LN, Eq.~\ref{eq:LN}). 
\begin{subequations}
\begin{equation}
\label{eq:PL}
  \frac{dN}{dL} \propto 
  L^{-\alpha}\qquad
  L \leq L_\mathrm{max}
\end{equation}
\begin{equation}
\label{eq:PLexp}
  \frac{dN}{dL} \propto 
  L^{-\alpha}
  e^{-(L/L_c)^{-\beta}}
\end{equation}
\begin{equation}
\label{eq:BPL}
  \frac{dN}{dL} \propto 
  \left\{
  \begin{split}
      &L^{-\alpha_1}\qquad
      L \leq L_b\\
      &L^{-\alpha_2}\qquad
      L_b <L
  \end{split}
  \right.
\end{equation}
\begin{equation}
\label{eq:LN}
  \frac{dN}{dL} \propto \frac{1}{L} 
  	\exp\left[-\frac
    {\left(\log_{10} L - \log_{10} L_0\right)^2}
    {2\sigma_L^2}\right]
\end{equation}
\end{subequations}
Unless specified, we fix the minimum and maximum luminosities to $L_\mathrm{min}=10^{30}\mathrm{\,erg\,s^{-1}}$ and
$L_\mathrm{max}=10^{37}\mathrm{\,erg\,s^{-1}}$ respectively. The number of free parameters varies for different scans. For a single power law we have the slope ($\alpha$) and the hard-cutoff ($L_\mathrm{max}$). The power law with super-exponential cutoff has the slope ($\alpha$), cutoff luminosity ($L_c$) and $\beta$. For a broken power law we have the low and high luminosity slope along with the break luminosity, denoted $\alpha_1$, $\alpha_2$, and $L_b$ respectively. Finally, for the log-normal distribution we have the peak of the distribution and its width denoted $L_0$ and $\sigma_{L}$ respectively. All parameters and their prior ranges
are given Table.~\ref{tab:params}.
The probability distributions for the luminosities
are directly proportional to the luminosity function $P(L) \propto dN/dL$, with $\int_{L_{\text{min}}}^{L_{\text{max}}} P(L) dL = 1$.

\subsubsection{Spatial profiles}
\label{sec:rho}
We consider two different functional forms 
for the disk. Each density profile is defined
in cylindrical coordinates ($r, z, \theta$) centered on the Galactic center.
The probability of finding a source at a given location
is proportional to the density profile
$P(r, z, \theta) = r\, n(r, z, \theta) / N$. Using the appropriate
coordinate transformation (see Appendix \ref{sec:coordinates}) 
this probability can be transformed to the probability
of finding a source at galactic longitude and latitude $(\ell, b)$ and at distance $D$ from the Sun:
$P(\ell, b, D) = D^2\cos{\left(b\right)} n(r, z, \theta) / N$.

Below we discuss the two parameterizations of the disk profile considered in this work,
our benchmark is the Lorimer profile \citep{Lorimer:2006qs}.
In addition, we also test a model with a gaussian radial profile
\citep{FaucherGiguere:2009df}.

\paragraph{Lorimer-disk profile}
The Lorimer profile has a radial distribution that is described
by a gamma function, whereas the $z$ distribution follows an exponential.
The number density of sources is given by \citep{Lorimer:2006qs}:
\begin{equation}
\label{eq:rho_lor}
  \begin{split}
  n\left(r, z\right) =& 
  N \frac{C^{B+2}}
  {4 \pi R^2_\odot z_s e^C \Gamma\left(B+2\right)} \times \\
  &\left(\frac{r}{R_\odot}\right)^{B} 
  \exp\left[-C\left(\frac{r-R_\odot}{R_\odot}\right)\right] \times\\
  &\exp\left(-\frac{|z|}{z_s}\right).
  \end{split}
\end{equation}
Here $N$ is the number of disk sources, 
$\Gamma$ the 
gamma function, $B$ and $C$ are parameters that define the spatial 
radial profile, $z_s$ is scale height and $R_\mathrm{\odot}=8.5\mathrm{\,kpc}$
the Solar distance from the Galactic Center. 
The spatial parameters $B, C$ and $z_s$ are left free in the scan (see Table \ref{tab:params}).
We note that the Lorimer disk reduces to a spatial profile with an exponential 
radial profile as considered by \cite{Story:2007xy} for $B=0$.

\paragraph{Gaussian radial profile}
We also consider a spatial profile with an exponential disk and a Gaussian
radial profile \citep{FaucherGiguere:2009df}:
\begin{equation}
  \label{eq:rho_gauss}
  n\left(r, z\right) = 
  N \frac{1}{4 \pi \sigma_r^2 z_s} e^{-r^2/2\sigma_r^2}e^{-|z|/z_s}.
\end{equation}

\paragraph{Bulge profile}
Motivated by the GCE, we allow for the presence of a bulge population of MSPs
in addition to the disk population in a subset of our scans.
We model the bulge as a radial power-law with a hard cutoff at $r_c = 3\mathrm{\,kpc}$
and fixed slope of $\Gamma=2.5$ \citep{Calore:2014xka, Daylan:2014rsa},
\begin{equation}
\label{eq:rho_bulge}
  n(r) = N_\mathrm{bulge} \frac{3-\Gamma}{4\pi r_c^{3-\Gamma}}r^{-\Gamma}.
\end{equation}
Again, $P(\ell, b, D) = D^2\cos\left({b}\right) n(r, \theta, \phi) / N_{bulge}$ 
(see Appendix \ref{sec:coordinates}).

Recently, it was found that the GCE is better described by a morphology that
traces the triaxial boxy bulge instead of a spherically-symmetric profile
\citep{Bartels:2017vsx, Macias:2016nev}.
Nevertheless, we model the bulge MSP population with a radial power-law.
The goal is to test whether this component is required by the data at all.
We do not expect this analysis to be sensitive to the exact morphology 
of the bulge.

\subsubsection{Detection sensitivity}
\label{sec:Fth}
We allow for some uncertainty in the Fermi detection sensitivity.
Depending on the dataset we use, the true detection efficiency can be an arbitrarily 
complicated function. 
In particular for confirmed pulsars, many of which have been detected 
by folding in the radio pulsation period,
it does not only depend on the $\gamma$-ray brightness of a source,
but also on the radio properties of the pulsar population and the sensitivity
of current radio telescopes. Therefore, we expect the sensitivity to be different
from the Fermi detection sensitivity.
Here we follow the same procedure as \citep{Hooper:2015jlu,Ploeg:2017vai} to model the detection sensitivity.
The threshold flux at a given sky position is drawn from a log-normal distribution:
\begin{equation}
\begin{split}
    P&(F_\mathrm{th}|\ell, b) = \frac{1}{\sigma_\mathrm{th} F_\mathrm{th} \sqrt{2\pi}} \\
    & \exp\left[-\frac{\left(\ln F_\mathrm{th} - \left(\ln(F_\mathrm{th,\,mod.}(\ell,b)) + K_\mathrm{th}\right)\right)^2}{2\sigma_\mathrm{th}^2}\right]
    \;,
\end{split}
\end{equation}
where $F_\mathrm{th,\,mod.}(\ell,b)$ is the sensitivity map in Fig.~16 of \cite{TheFermi-LAT:2013ssa}. We have two free parameters $K_\mathrm{th}$ and $\sigma_\mathrm{th}$, respectively the normalization and width of the distribution from which $F_\mathrm{th}$ is drawn.
A source is detected if $F\geq F_\mathrm{th}$, therefore
\begin{multline}
	\label{eq:Pth}
    P_\mathrm{th}(F|\ell, b) \equiv 
    	P\left(F\geq F_\mathrm{th}|\ell, b\right)\\ = \frac 1 2 + 
        \frac 1 2 \mathrm{erf}\left[\frac{\ln F - 
    	\left(\ln(F_\mathrm{th,\,mod.}(\ell,b)) + K_\mathrm{th}\right)}
    	{\sqrt{2}\sigma_\mathrm{th}}\right] \;.
\end{multline}
%
\subsubsection{Flux uncertainties}
\label{sec:flux}
Energy fluxes ($0.1-100\mathrm{\,GeV}$) and their uncertainties are taken from the 
2FGL \citep{Fermi-LAT:2011yjw}, 3FGL \citep{Acero:2015hja}
or the preliminary Fermi--Lat 8-year catalog 
(FL8Y)\footnote{\url{https://fermi.gsfc.nasa.gov/ssc/data/access/lat/fl8y/}}.
The flux uncertainties are treated as Gaussian, the probability
of a source having some true flux ($F$) is given by
\begin{equation}
  P(F|F_\mathrm{obs}) = \frac{1}{\sqrt{2\pi \sigma_{F}^2}}
  e^{-\left(F - F_\mathrm{obs}\right)^2 / 2 \sigma_{F}^2},
\end{equation}
where $F_\mathrm{obs}$ and $\sigma_F$ are the observed
energy flux ($\geq0.1\mathrm{\,GeV}$) and its associated uncertainty.

\begin{table}[h!]
\centering
\label{tab:params}
\begin{tabular}{l|cc}
\tablewidth{0pt}
\hline
\hline
Parameter & prior & fixed \\
\hline
$\log_{10}{N}$					&	$\left[0, 8\right]$		& - \\
$\log_{10}{N_\mathrm{bulge}}$	&	$\left[0, 8\right]$		& - \\
\hline
\multicolumn{3}{c}{Luminosity function}\\
\hline
$\log_{10}L_\mathrm{min}$	&	- 										& $30$ \\
$\log_{10}L_\mathrm{max}$			&	
$(\left[33.7, 37\right])$ 	& 
$37$ \\
$\log_{10}{L_c}$			&	$\left[32, 37\right]$		& - \\	
$\log_{10}{L_b}$			&	$\left[31, 37\right]$		& - \\	
$\log_{10}{L_0}$			&	$\left[31, 37\right]$		& - \\	
$\alpha, \alpha_1, \alpha_2$	&	$\left[0.1, 5.0\right]$ 						& - \\
$\sigma_L$			&	$\left[0.5, 5\right]$		& - \\	
$\beta$			&	$\left[0, 3\right]$		& - \\	
\hline
\multicolumn{3}{c}{Spatial profile}\\
\hline
$B$		&	$\left[0, 10\right]$		&	-	\\
$C$		&	$\left[0.05, 15\right]$	&	-	\\
$z_s$	&	$\left[0.05, 3\right]$		&	-	\\
$\sigma_r$	&	$\left[0.05, 15\right]$		&	-	\\
$r_c$		& - &	$3$		\\
$\Gamma$	& - &	$2.5$	\\
\hline
\multicolumn{3}{c}{Detection sensitivity}\\
\hline
$\sigma_\mathrm{th}$	&	$\left[0.05, 3\right]$		&	-	\\
$K_\mathrm{th}$			&	$\left[-3, 3\right]$		&	-	\\
\hline
\end{tabular}
\caption{All parameters of the likelihood with their prior values or 
the value they are fixed too. $L_\mathrm{max}$ is only left
free in case a single power law with hard cutoff is fitted for.}
\end{table}

\subsection{\label{sec:dist}Distances}
There are two primary methods for measuring the distances to pulsars. If they are close enough to our galactic position it can be possible to obtain a parallax distance measure, typically accepted as the most unbiased method to measure distances to pulsars. However, for the majority of pulsars the only distance measure comes from radio observations of the dispersion measure (DM), a frequency dependent time shift of the pulse profile. 
In order to take into account uncertainties in the distance estimates we construct
a realistic probability-density function (PDF) for the probability of measuring
a specific parallax ($w_\mathrm{obs}$) or dispersion measure ($\mathrm{DM}_\mathrm{obs}$) given a true distance
to the source: $P\left(\kappa_\mathrm{obs}|D\right)$ with $\kappa_\mathrm{obs}$ being the parallax or dispersion measure. In the likelihood we then integrate over $D$. 
If parallax information is available we construct distance PDFs using these measurements, otherwise we use DM information.
In case neither is available, this term is not present in the likelihood (Eq.~\ref{eq:prob_nodist}).
\subsubsection{Distance from Parallax}
For a small number of MSPs in our sample parallax information is available (see Tab.~\ref{tab:sources}). 
True parallaxes ($\omega(D) \equiv 1 / D$) and
measured uncertainties ($\sigma_{\omega_\pm}$) 
are used to construct a PDF for the observed parallax $\omega_\mathrm{obs}$. 
The error on the parallax is taken to be Gaussian, but can be asymmetric. The PDF for the distance can then be constructed as follows \citep{Verbiest:2012kh}
\begin{multline}
    P\left(\omega_\mathrm{obs}|\omega(D)\right) \propto   \\
    \Theta_H\left(\frac 1 D - \omega_\mathrm{obs}\right)
    \exp\left[-\frac 1 2 \left(\frac{\omega_\mathrm{obs} - 1/D}
    {\sigma_{\omega_+}}\right)^2\right]  \\
   +\Theta_H\left(\omega_\mathrm{obs} - \frac 1 D\right)
    \exp\left[-\frac 1 2 \left(\frac{\omega_\mathrm{obs} - 1/D}
    {\sigma_{\omega_-}}\right)^2\right],
\end{multline}
where $\Theta_H$ is the heaviside-step function.

\subsubsection{Distance from DM}
\label{sec:dist_dm}
\begin{figure}[h!]
  \centering
  \includegraphics[width=\linewidth]{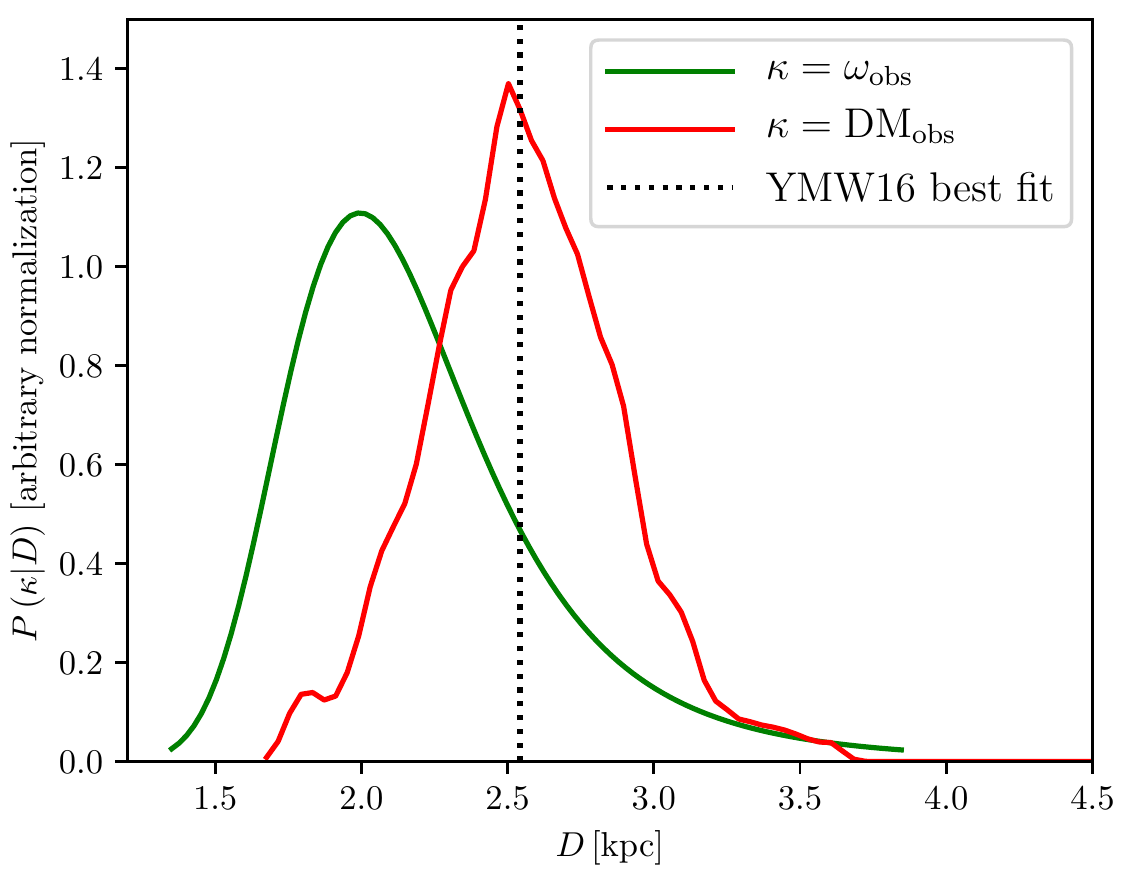}
  \caption{
  Probability distribution for the measured parallax and dispersion
  measure of J1600-3053 given the true distance of the source.
  PDFs have an arbitrary normalization.
  The black-dotted line shows the distance corresponding the observed dispersion measure
  and the best-fit parameters of the YMW model.
  Varying the parameters of the YMW16 model yields the distribution shown in red.
  The green line corresponds to the PDF for the parallax.
  }
  \label{fig:MSPPDF}
\end{figure}

The origin of the DM is assumed to come from interactions with free electrons along the line-of-sight. Assuming a particular distribution of free electrons in the Galaxy we can therefore calculate the distance to any given pulsar using,
\begin{equation}
\text{DM} = \int^D_0 n_e(l)\, \text{d}l\,,
\end{equation}
where $n_e$ is number density of electrons along the line-of-sight.
Whereas the DM for each source is well constrained, $n_e(l)$ is a source of large uncertainties for individual sources \citep{Lorimer:2001vd} and sometimes the cause of systematic biases \citep{2017ApJ...835...29Y}. To date there are three main models for $n_e$: TC93 \citep{Taylor:1993my}, NE2001 \citep[][used by the majority of past MSP luminosity function analyses]{Cordes:2002wz}, and the recent YMW16 \citep{2017ApJ...835...29Y}. \cite{2017ApJ...835...29Y} showed that the YMW16 model was less affected by the large errors which typically entered the NE2001 model, particularly at high galactic latitudes, the regime in which NE2001 was shown to have large systematic biases \citep{2011AIPC.1357..127R}. We assume the YMW16 model as a description of the electron density. The YMW16 model contains 35 free parameters describing a variety of galactic components contributing to the total electron density, for example the scale height of the thick disk. In principle these could all affect the distance calculated to a given source. 
For each pulsar a PDF is generated for the observed dispersion measure as a function of true distance.
We adopt a conservative approach by sampling from all variable parameters and calculating
the dispersion measure for each pulsar given a true distance to the source.
Gaussian distributions around each parameter are assumed with the central values and $1\sigma$ errors as provided in Table 2 of \cite{2017ApJ...835...29Y}. 
We sample $10^5$ combinations of parameters and true distances for each pulsar and 
create a PDF by binning the data in a histogram.
An example is provided in Fig.~\ref{fig:MSPPDF}. Using this method, we found that the PDF always peaks extremely close to the best fit value from the YMW16 model but there can be quite significant spread, even though most of the parameters in the YMW16 model are quite well constrained.

All code to reproduce the DM-based probability-distribution functions for
either the dispersion measure or
the distance to an individual pulsar are publicly available at \url{https://github.com/tedwards2412/MSPDist}. 
We provide a python wrapper for the YMW16 electron-density model
\citep{2017ApJ...835...29Y} and accompanying code to calculate distance
uncertainties. 

\subsection{Pulsar sample}
\subsubsection{$\gamma$-ray detected pulsars}
In our benchmark analysis we exclusively use the $\gamma$-ray detected 
MSPs not associated
with a globular cluster. All sources have spin periods $\leq 30\mathrm{\,ms}$.
Our sample contains 96 sources with confirmed $\gamma$-ray pulsations
(see Tab.~\ref{tab:sources}). 
The source list is compiled using the second pulsar catalog (2PC) \citep{TheFermi-LAT:2013ssa}
and the public list of \Fermi-LAT detected $\gamma$-ray 
pulsars as was available on May 14 2018
\footnote{\url{https://confluence.slac.stanford.edu/display/GLAMCOG/ Public+List+of+LAT-Detected+Gamma-Ray+Pulsars}}.
Unless specified otherwise, fluxes are taken from the third \Fermi-LAT source catalog 
\citep[3FGL, ][]{Acero:2015hja}. 
When a pulsar is not present in the 3FGL we also look for fluxes in 
the second \Fermi-LAT source catalog \citep[2FGL, ][]{Fermi-LAT:2011yjw} 
and the FL8Y.
Similarly, parallax and dispersion measures are obtained from the ATNF catalog
\citep{Manchester:2004bp}.

\subsubsection{Unassociated sources}
The 3FGL contains $3033$ objects with roughly a third still unassociated to a particular source type. Follow-up radio observations of many of the unassociated sources have shown that there could be a large population of pulsars still remaining to be found within the 3FGL. If only a small proportion turn out to be MSPs this population will still tend to dominate the overall data set. 
We therefore must attempt to take this population into account and see how it
could systematically affect our results.
We capture the possible effects of the unassociated sources by presenting three scenarios. 
First, we perform our analysis using only the 96 $\gamma$-ray detected sources.
In addition, we perform the same analysis using only the 39 MSPs present in the 2PC.
Finally, we combine the 96 $\gamma$-ray detected sources with 69 sources without $\gamma$-ray detected pulsations based on the 
results from \cite{Parkinson:2016oab}. Although some of these 69 sources have
unconfirmed associations, 
we will refer to this sample as unassociated sources for conciseness.
These can be found in Table \ref{tab:sources} under 'other sources'.

\cite{Parkinson:2016oab} performed a classification analysis of the 3FGL using a variety of Machine Learning tools, the most accurate being Random Forest which achieved $>90\%$ correct associations when trained on $70\%$ of the sample and tested on the remaining $30\%$.
For the construction of our unassociated sample, we select all 3FGL unassociated sources
and source candidates of any given class that have not been confirmed. We 
require that each source is classified as a pulsar by either the logistic regression 
or Random Forest analysis of \cite{Parkinson:2016oab} with over 50\% probability. 
Moreover, we require the same classifier to classify the candidate as an MSP rather than a
young pulsar. Finally, we require a detection significance in the 3FGL or FL8Y of
$\geq 10\sigma$, similar to the list in Table 6 of \cite{Parkinson:2016oab} to optimize
the chances of the classification being correct. We note a few of the prime candidates
in this table have since been discovered as $\gamma$-ray MSPs, including the two recent detections
by \cite{Clark:2018zsp}.

\subsection{Parameter scan}
We efficiently scan the parameter space using the Bayesian nested sampling package \texttt{MultiNest} \citep{Feroz:2008xx, Buchner:2014nha}. For the low-dimensional problems at hand, \texttt{MultiNest} is accurate and requires a computationally feasible number of likelihood calculations to accurately map the posterior distribution. 
In addition it is able to handle multi-modal distributions and degeneracies in the parameter space, the latter being a problem we are likely to encounter when considering particular configurations of luminosity functions, such as PL with a maximum luminosity cut-off.
The results presented in Sec.~\ref{sec:results} use \texttt{nlive}$\,=500$.

For each model the Bayesian evidence is computed \citep[e.g.~][]{Trotta:2008qt}
\begin{equation}
\label{eq:ev}
  \mathcal{Z} = P\left(\mathcal{D}\right) = \int  \mathcal{L}\left(\mathbf{\Theta}\right)
  \mathcal{\pi}(\mathbf{\Theta})d\mathbf{\Theta},
\end{equation}
where $\mathcal{\pi}(\Theta)$ is the prior on each parameter.
The Bayes factor is then defined 
as
\begin{equation}
\label{eq:bayes}
\begin{split}
  B_{12} &\equiv \frac{P\left(H_2\right|\mathcal{D})}{P\left(H_1\right|\mathcal{D})} 
  = \frac{\mathcal{Z}_2 P(H_2)} {\mathcal{Z}_1 P(H_1)}, 
\end{split}
\end{equation}
\begin{sloppypar}
with $H_{1,2}$ denoting the different models \citep{Trotta:2008qt}. 
We choose equal priors for different models,
$P(H_2)/P(H_1) = 1$. Since our models are not nested hypotheses, Bayesian model selection, which does not require this assumption, provides a straightforward comparison of our models. We note that, in contrast to Frequentist analyses, it is here relevant to properly normalize the likelihood functions in order to make the evidence and the Bayes factor informative. The expressions in Eq.~\ref{eq:likelihood} and Eq.~\ref{eq:prob} ensure this.
\end{sloppypar}

\section{\label{sec:results}Results}

\subsection{Model comparison}
For each of the three data sets ($\gamma$-ray detected 
MSPs, MSPs plus MSP candidates from \cite{Parkinson:2016oab} and the 2PC MSPs)
we compare multiple models, each characterized
by their luminosity function, spatial profile and whether or not we included a bulge population.

In order to interpret the results we use Bayesian model comparison following 
\cite{10.2307/2291091}. We compute
$2\ln B_{12}$ from Eq.~\ref{eq:bayes} always comparing against a benchmark model 
($H_2$: BPL, Lorimer).
If $2\ln B_{12} \in [0, 2]$ there is no preference for $H_2$ over $H_1$.
$2\ln B_{12} > 10$ represents strong preference for $H_2$. Contrarily,
$2\ln B_{12}<0$ indicates $H_1$ is preferred over $H_2$.

\begin{table}[h]
    \centering
    \begin{tabular}{lcc}
      \toprule
      Model & $\ln\mathcal{Z}$ & $2\ln B_{12}$\\
      \midrule
\multicolumn{3}{c}{$\gamma$-ray detected pulsars}\\
\hline
                   BPL, Lorimer & 2042.0 & 0.0\\
            BPL, Lorimer, bulge & 2041.6 & 0.8\\
                    LN, Lorimer & 2040.0 & 4.0\\
             LN, Lorimer, bulge & 2040.0 & 4.0\\
        PL exp.~cutoff, Lorimer & 2036.6 & 10.8\\
                  BPL, gaussian & 2024.0 & 36.0\\
           BPL, gaussian, bulge & 2023.7 & 36.6\\
                   LN, gaussian & 2021.8 & 40.4\\
            LN, gaussian, bulge & 2021.2 & 41.6\\
                    PL, Lorimer & 2017.6 & 48.8\\
\hline
\multicolumn{3}{c}{All sources}\\
\hline

                   BPL, Lorimer & 3889.6 & 0.0\\
            BPL, Lorimer, bulge & 3889.6 & 0.0\\
                    LN, Lorimer & 3888.3 & 2.6\\
                  BPL, gaussian & 3875.9 & 27.4\\
                   LN, gaussian & 3874.4 & 30.4\\
\hline
\multicolumn{3}{c}{2PC}\\
\hline

                   BPL, Lorimer & 789.0 & 0.0\\
                    LN, Lorimer & 787.9 & 2.2\\
                  BPL, gaussian & 780.0 & 18.0\\
                   LN, gaussian & 778.8 & 20.4\\
      \bottomrule
      \end{tabular}
    \caption{Model comparison for the three different 
    datasets analyzed. Each model is characterized by the luminosity function,
    spatial profile and whether or not we included a bulge population.
    We show the log of the Bayesian evidence ($\ln\mathcal{Z}$)
    for each model and the Bayes factor ($B_{12} = 2 \ln \mathcal{Z}_{2} / \mathcal{Z}_{1}$)
    with respect to the best-fitting model without bulge
    \citep{10.2307/2291091}.}
    \label{tab:lg_be}
\end{table}

The results for the various \texttt{MultiNest} scans performed are shown in
Tab.~\ref{tab:lg_be}. Each dataset is shown separately and models are ordered by
decreasing $\mathcal{Z}$.
Our default dataset ($\gamma$-ray detected pulsars only) shows that a single power-law parameterization of the luminosity
function,
regardless of whether it has a hard or super-exponential cutoff, is greatly disfavored.
No strong preference is present for either a log-normal or broken power-law parameterization,
although the latter performs slightly better.
Concerning the spatial profile, the Lorimer disk is strongly preferred over the radial
Gaussian profile. 
No bulge component is required by the data. 
A small point of caution, in a few cases the evidence of models including the bulge
is smaller than of identical models without a bulge component.
However, the likelihood for the models including the bulge is higher than that of
those where it is not included, which is expected
when including additional degrees of freedom. The fact that the evidence goes down
with the addition of a new component means that the model without the additional
component suffices to describe the data.
Given these results, we will henceforth consider the Lorimer disk with a BPL luminosity function
and no bulge as our benchmark model and show results for this run. Additional results can be found in Appendix \ref{sec:other_results}.

\subsection{Parameters}
\begin{figure*}[t]
  \centering
  \includegraphics[width=0.99\linewidth]{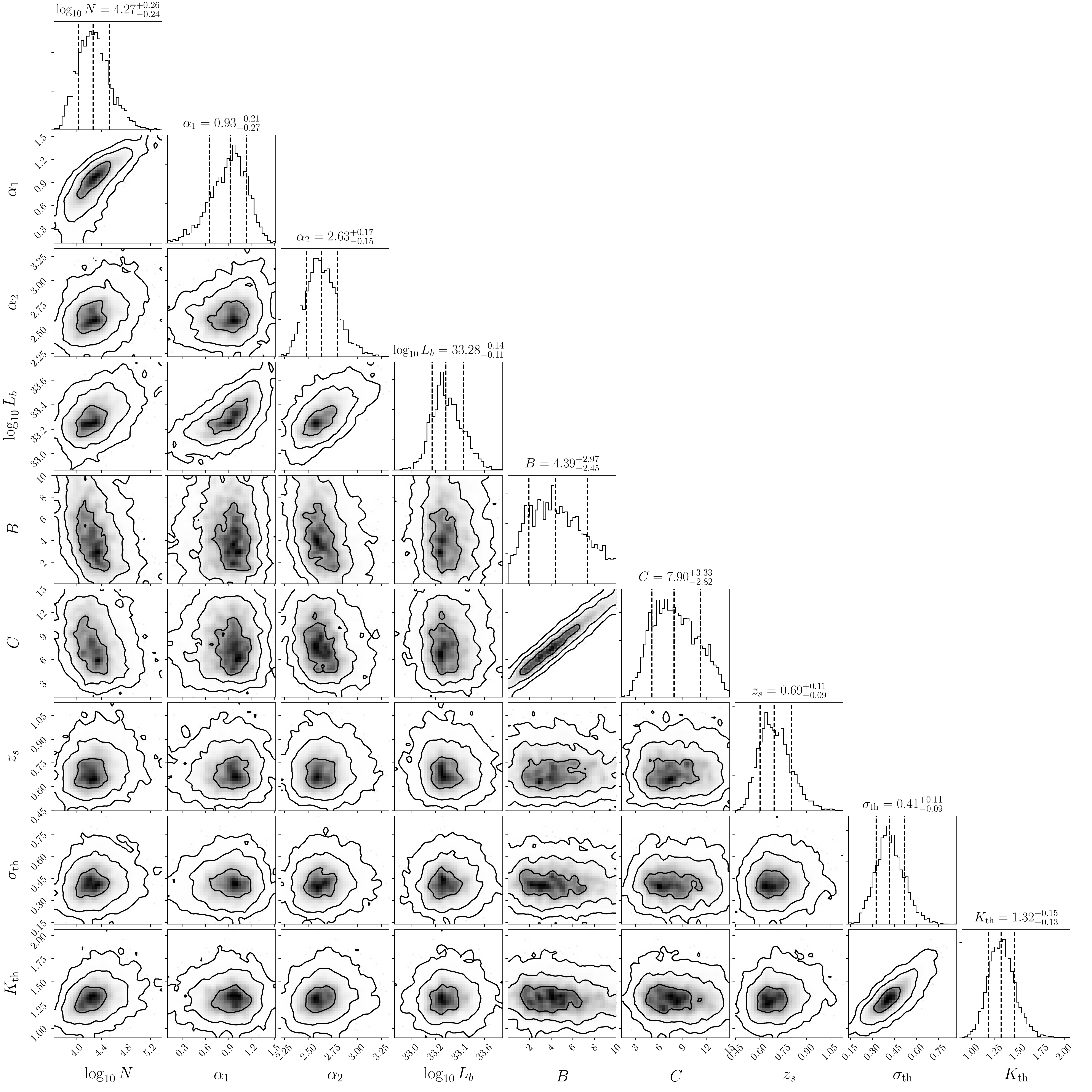}
  \caption{Corner plot for the parameters of our benchmark model.
  Contours in the two-dimensional histogram are $1, 2$ and $3\sigma$.
  Dashed-lines in the one-dimensional posterior show the
  $16, 50$ and $84\%$ quantiles. Values above each posterior
  represent the $50\%$ quantile with $1\sigma$ errors.}
  \label{fig:bench}
\end{figure*}

In Fig.~\ref{fig:bench} we show a corner plot for the parameters of our benchmark model.
Contours in the two-dimensional histograms are $1, 2$ and $3\sigma$.
Dashed-lines in the one-dimensional posterior represent $16, 50$ and $84\%$
quantiles.
The best fit parameters for our benchmark model and for the log-normal luminosity
function with a Lorimer disk are given in Tab.~\ref{tab:bestfit}.
Corner plots for other representative models in Tab.~\ref{tab:lg_be} are
presented in Appendix \ref{sec:other_results}.
\begin{table}[h]
\centering
\caption{Best-fit parameters and characteristics 
for the populations with a broken power-law and log-normal
luminosity function and Lorimer-disk spatial profile.
Luminosities and fluxes are in the range $0.1-100\mathrm{\,GeV}$.}
\label{tab:bestfit}
\begin{tabular}{lcc}
\tablewidth{0pt}
\toprule
Parameter & broken power-law & log-normal \\
\midrule
\multicolumn{3}{c}{Luminosity function}\\
\hline
$\log_{10}L_\mathrm{min}$		&	$30$ 		& $30$ 		\\
$\log_{10}L_\mathrm{max}$		&	$37$		& $37$ 		\\
$\alpha_1$						& 	$0.97$		& - 		\\
$\alpha_2$						& 	$2.60$		& - 		\\
$\log_{10}{L_b}$				&	$33.24$		& - 		\\	
$\log_{10}{L_0}$				&	-			&	$32.61$	\\	
$\sigma_L$						&	-			&	$0.63$	\\
\hline
\multicolumn{3}{c}{Spatial profile}\\
\hline
$B$								&	$3.91$		&	$2.75$	\\	
$C$								&	$7.54$		& 	$5.94$	\\
$z_s$							& 	$0.76$		&	$0.63$	\\
\hline
\multicolumn{3}{c}{Detection sensitivity}\\
\hline
$\sigma_\mathrm{th}$			& 	$0.41$		&	$0.45$	\\
$K_\mathrm{th}$ 				& 	$1.35$		& 	$1.33$	\\
\hline
\multicolumn{3}{c}{Other characteristics}\\
\hline
$\log_{10}{N}$										&	$4.38$ 				&	$4.12$ 		\\
$\left<L\right> \mathrm{\,[erg\,s^{-1}]}$			&	$6.2\times10^{32}$	&   $1.1\times10^{33}$	\\
$L_\mathrm{tot} \mathrm{\,[erg\,s^{-1}]}$			&	$1.5\times10^{37}$	&	$1.5\times10^{37}$	\\
$F_\mathrm{tot} \mathrm{\,[erg\,cm^{-2}\,s^{-1}]}$	&	$4.7\times10^{-9}$	&	$4.8\times10^{-9}$	\\
Expected bulge detections							&	$4.5$				&	$2.9$ \\
\bottomrule
\end{tabular}
\end{table}

The total number of sources with $L\geq L_\mathrm{min}$ is $\sim 2\times10^{4}$ 
for our best-fit model. However, it could be as small as $\sim10^{4}$ 
or as large as $\sim 10^{5}$. Unlike previous claims 
\citep{Gregoire:2013yta}, we find the $\gamma$-ray MSP population to be 
compatible with the
the expected number of MSPs from population studies using radio pulsars \citep{Cordes:1997my, 1998MNRAS.295..743L, Levin:2013usa}.

\begin{figure}[t]
  \centering
  \includegraphics[width=\linewidth]{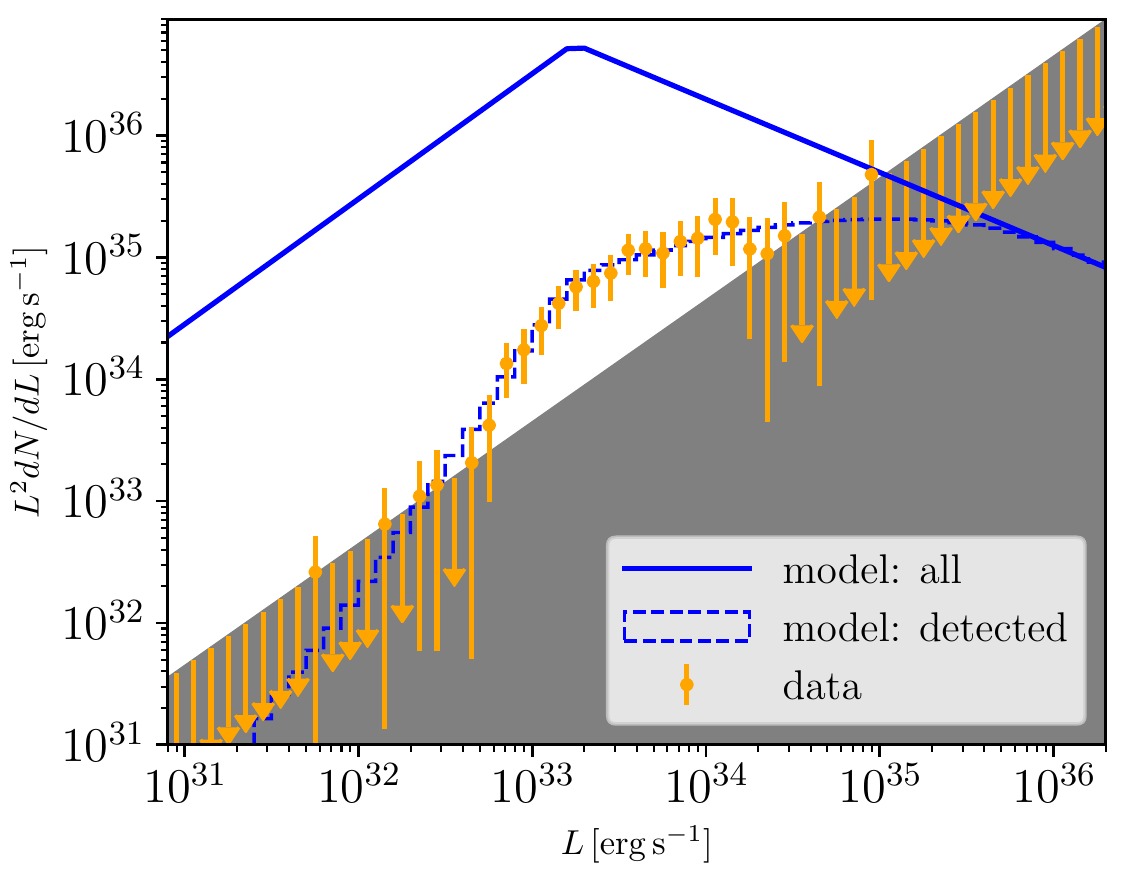}
  \caption{Luminosity function ($0.1-100\mathrm{\,GeV}$) of our benchmark model.
  The blue solid line shows the total luminosity function, whereas the 
  dashed line only shows the expected sources. Orange errorbars are the
  expectations-values from the data where distance and 
  flux uncertainties have been taken into 
  account (for more details see Appendix
  \ref{sec:L_err}).
  The grey-shaded area corresponds to one or fewer sources.}
  \label{fig:dNdL_bench}
\end{figure}

\paragraph{Luminosity function}
In Fig.~\ref{fig:dNdL_bench} we show the luminosity function. 
The blue solid line displays the total luminosity function,
whereas the dashed line shows the luminosity function with the detection efficiency
folded in. The grey shaded area corresponds to one or fewer sources at this luminosity.

\begin{figure}[t]
  \centering
  \includegraphics[width=\linewidth]{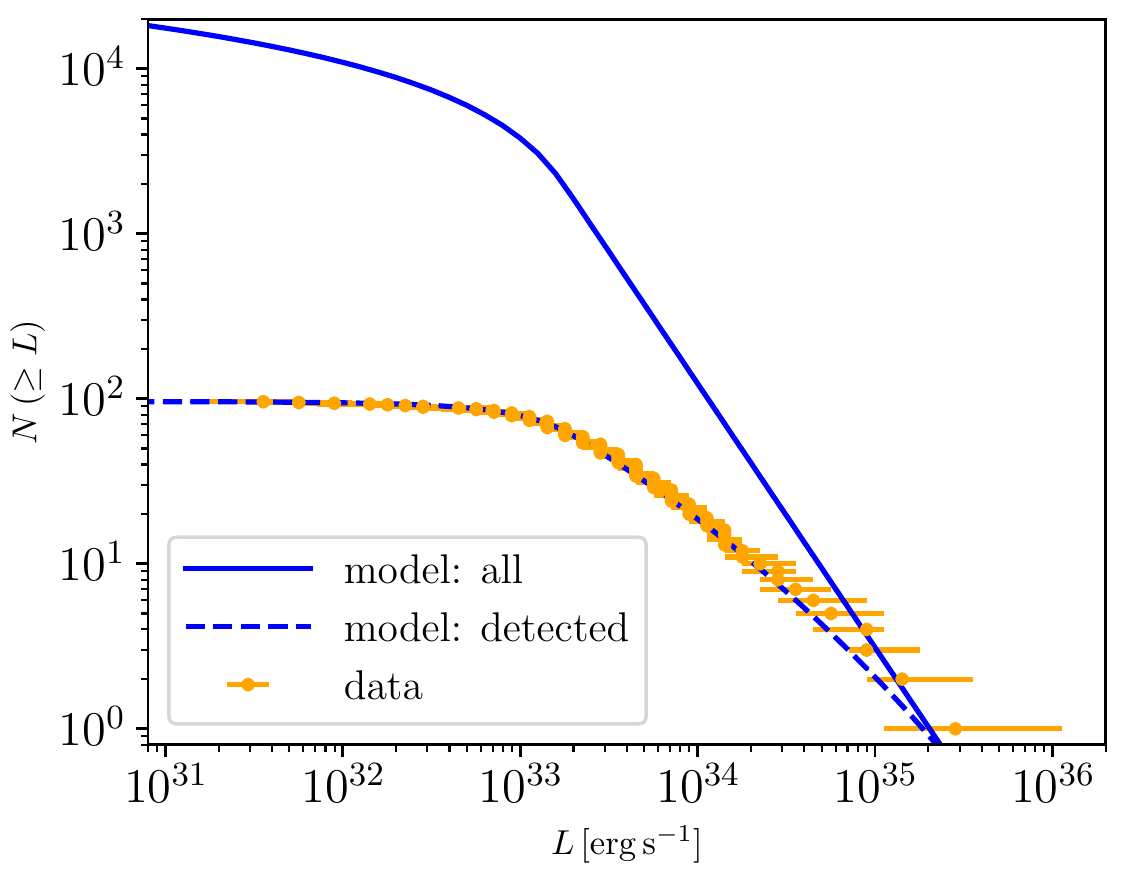}
  \caption{Similar to Fig.~\ref{fig:dNdL_bench}, but showing the cumulative distribution.
  Distance and flux uncertainties for individual pulsars are included in the errorbars,
  which show the median and 95\% containment interval.}
  \label{fig:dNdL_cum}
\end{figure}

Orange errorbars show the expectation values derived from the data. 
Uncertainties in the flux and distance to individual pulsars have been taken
into account (see Appendix \ref{sec:L_err}). 
Upper limits correspond to an expectation of fewer than one source in the particular bin.
In addition, we show the cumulative distribution of the luminosity function
in Fig.~\ref{fig:dNdL_cum}. 
The data point and errorbars show the median and the 95\% containment interval.

At $\sim 2\times 10^{33} \mathrm{\,erg\,s^{-1}}$ there is a clear turnover.
Due to the hard slope at low luminosities ($\alpha_1 = 1.0$) and soft slope at
high luminosities ($\alpha_2 = 2.6$) the total flux is dominated by sources
somewhat below the break luminosity. 
There is no indication of any MSPs brighter than $\mathrm{few}\times 10^{35}\mathrm{\,erg\,s^{-1}}$
or dimmer than $\sim 10^{32}\mathrm{\,erg\,s^{-1}}$.
This parameterization broadly agrees
with the results from \cite{Winter:2016wmy}.

\paragraph{Spatial profile}
Spatial parameters are not very well constrained. 
The scale height of the disk is $\sim 0.7{\rm\,kpc}$ but has an uncertainty
of a factor $\sim 1.5$, in broad agreement with earlier works
\citep[e.g.~][]{Story:2007xy, Levin:2013usa, Calore:2014oga,Hooper:2015jlu,Ploeg:2017vai}.
The radial parameters of the Lorimer profile are consistent with the distribution derived for the 
full radio pulsar population \citep{Lorimer:2003qc, Lorimer:2006qs} and with
expectations for the MSP population \citep{Lorimer:2015iga}. 
For the Gaussian profile (see Appendix \ref{sec:other_results}), the dispersion is $\sigma_r\sim 4\mathrm{\,kpc}$, but is again uncertain
by $\sim 25\%$. This result is consistent with the expectations
for an old pulsar population \cite{FaucherGiguere:2009df}. Our results for the
spatial profile are also in agreement with other analyses of $\gamma$-ray MSPs
\citep{Hooper:2015jlu,Ploeg:2017vai}.

\begin{figure}[t]
  \centering
  \includegraphics[width=\linewidth]{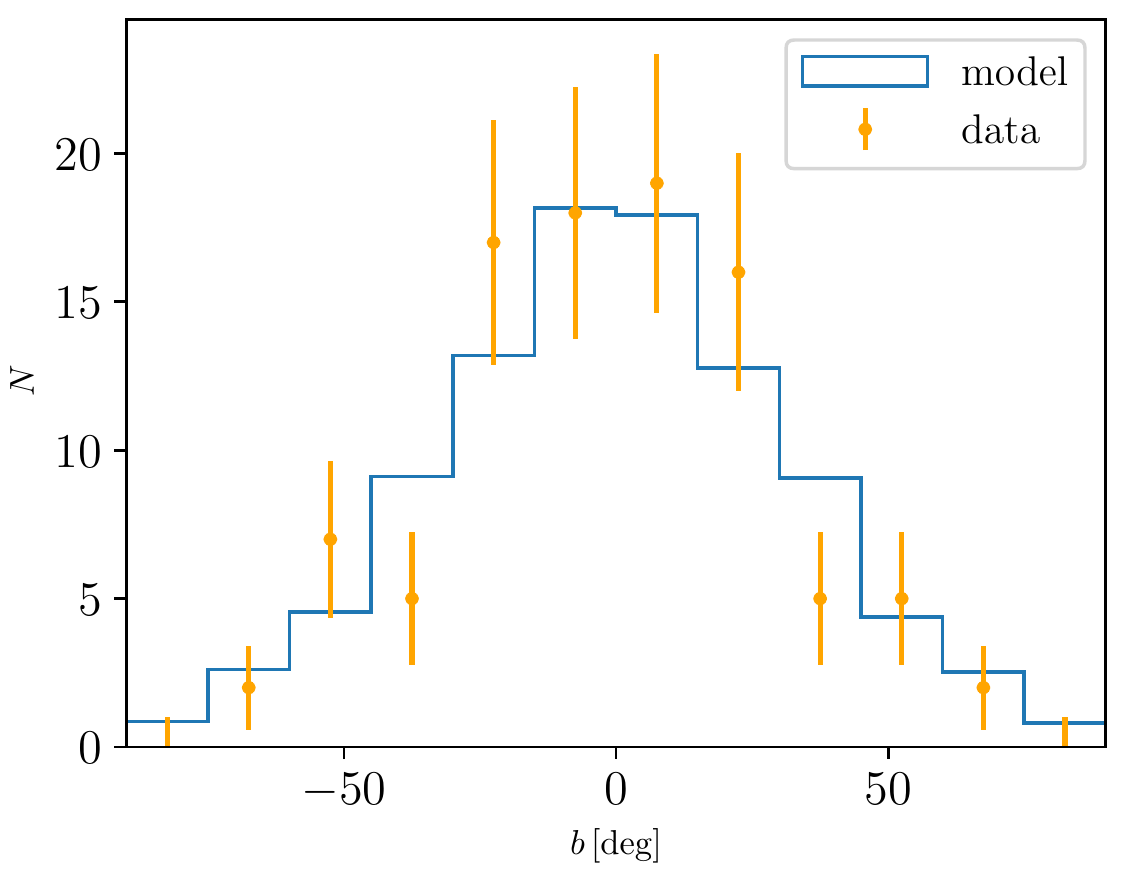}
  \caption{Latitude distribution of MSPs. Blue is the expected distribution. Orange
  data points show the observed distribution.}
  \label{fig:glat_bench}
\end{figure}
\begin{figure}[t]
  \centering
  \includegraphics[width=\linewidth]{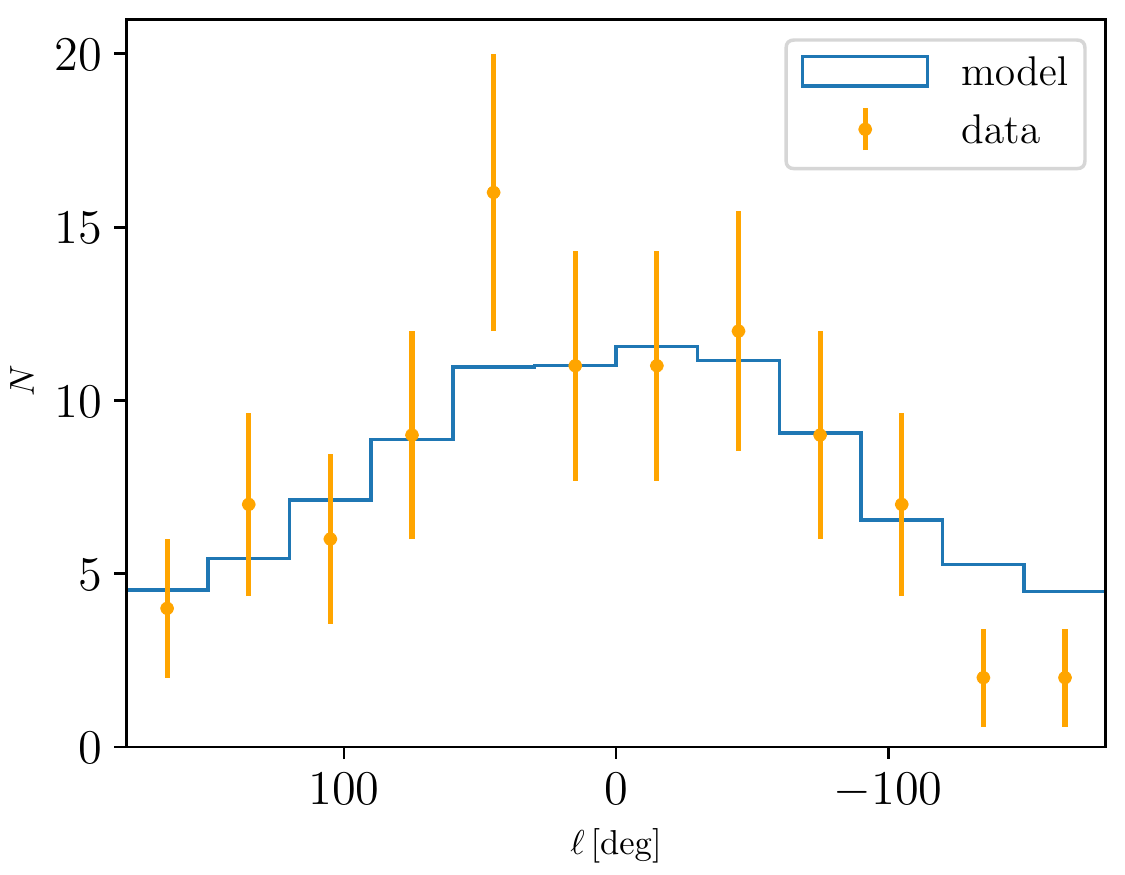}
  \caption{Same as Fig.~\ref{fig:glat_bench} but for longitude.}
  \label{fig:glon_bench}
\end{figure}
In Figs.~\ref{fig:glat_bench} and \ref{fig:glon_bench} we show the 
expected (blue) and observed (orange) latitude and longitude distribution of $\gamma$-ray detected MSPs.

\paragraph{Detection sensitivity}
In principle, the parameters $\left\{K_\mathrm{th}, \sigma_\mathrm{th}\right\}$ are
nuisance parameters. The positive value of $K_\mathrm{th}$ indicates
that our detection sensitivity is poorer than the sensitivity map we use
\citep{TheFermi-LAT:2013ssa}. However, this is not unexpected since \cite{TheFermi-LAT:2013ssa}
derived their map assuming $\gamma$-ray sources with a pulsar spectrum, but did not
require pulsations to be detected. We find, for the different datasets,
i.e.~2PC, $\gamma$-ray detected pulsars, and including not-yet-identified sources, the values
$\left\{K_\mathrm{th}, \sigma_\mathrm{th}\right\} = \left\{2.05, 0.64\right\}, 
\left\{1.35, 0.41\right\}$, and $\left\{1.19, 0.30\right\}$ respectively. Therefore, 
we see that the detection sensitivity improves with a larger sample which is expected
since a larger sample implies either increased exposure, such as when going from the 2PC to 
the full $\gamma$-ray detected pulsars sample, or a more lenient detection criteria, 
such as when we include unassociated sources.

\begin{figure}[t]
  \centering
  \includegraphics[width=\linewidth]{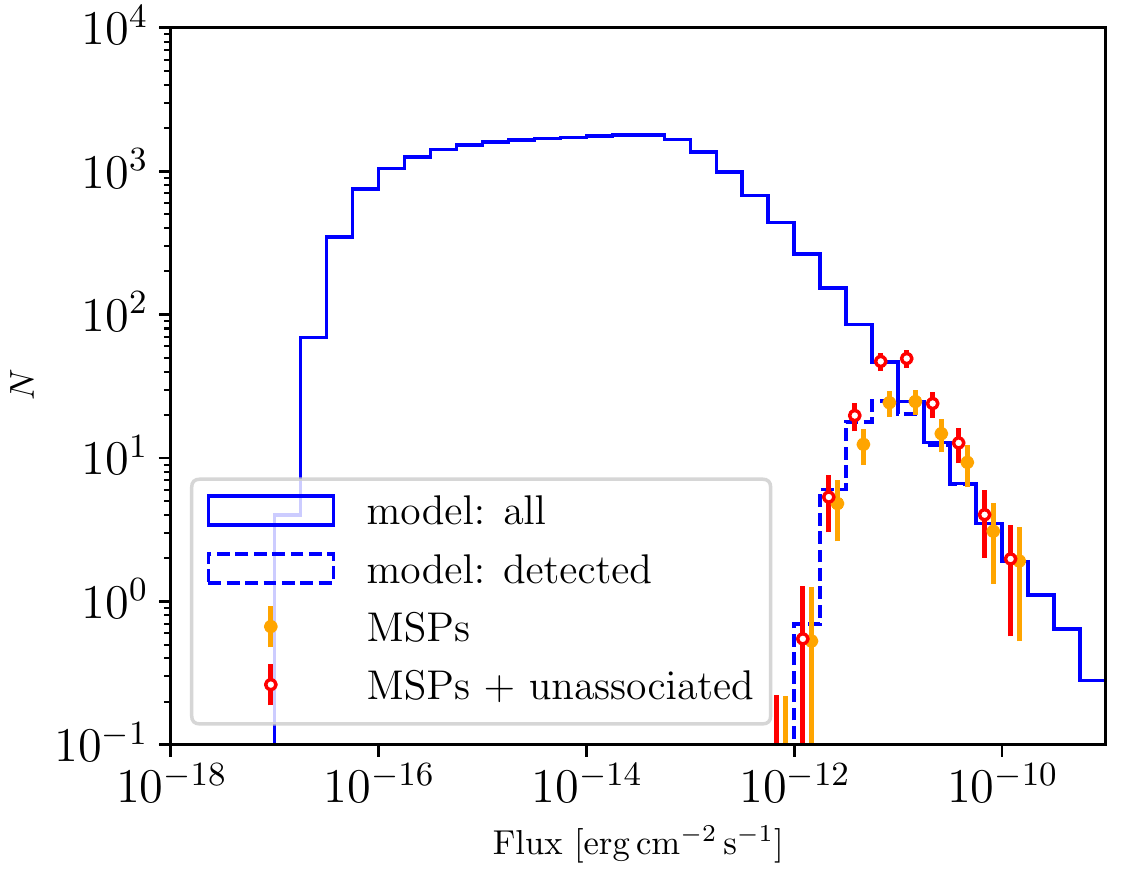}
  \caption{Flux distribution for our benchmark model. The blue solid line is
  the total MSP population. The dashed blue line takes into account the detection
  threshold. Orange errorbars are the data including all $\gamma$-ray detected pulsars.
  Red-open errorbars also include the 69 unassociated sources. Note that the blue-dashed
  line corresponds to the detection sensitivity derived using only the $\gamma$-ray detected
  pulsars.
  }
  \label{fig:flux}
\end{figure}

For completeness, we show the flux distribution in Fig.~\ref{fig:flux}.
The blue solid line is the total population, whereas the blue dashed line takes into
account the detection threshold. As can be seen our analysis suggests the MSP population is 
flux complete down to $F\gtrsim 10^{-11}\mathrm{\,erg\,cm^{-2}\,s^{-2}}$.

\subsection{Total Luminosity and Flux}
Given the number of sources and luminosity function we can determine the total luminosity.
Since the broken-power law peaks at luminosities of
$\sim 10^{33} \mathrm{\,erg\,s^{-1}}$ and has a hard (soft) slope at low (high) luminosities,
the total luminosity is fairly insensitive to $L_\mathrm{min}$ and $L_\mathrm{max}$.
The same holds for the log-normal distribution.
We find a total luminosity $L_\mathrm{tot} = 1.5\times 10^{37}\mathrm{\,erg\,s^{-1}}$
and a total flux of $4.7\times 10^{-9}\mathrm{\,erg\,cm^{-2}\,s^{-1}}$.
These numbers are uncertain by about a factor $\sim 2$.
Given a Milky-Way stellar-disk mass of 
$5.17\times 10^{10}\mathrm{\,M_\odot}$ \citep{Licquia:2014rsa}
we find that luminosity-per-stellar-mass for the Milky-Way disk is
$2.9\times10^{26}\mathrm{\,erg\,s^{-1}\,M_\odot^{-1}}$.


\section{\label{sec:discussion}Discussion}
\subsection{Unassociated sources}
Our default analysis includes 96 $\gamma$-ray detected MSPs. 
In addition, we performed analyses using only the 39 MSPs from the 2PC 
\citep{TheFermi-LAT:2013ssa} and
including an additional 69 unassociated sources
with selection criteria based on the results of \cite{Parkinson:2016oab}.
We find consistent results between the three analyses.
In particular, as can be seen in Tab.~\ref{tab:lg_be}, 
in all cases we find that there is no clear preference
for either a broken-power law or a log-normal luminosity-function parameterization. 
On the other hand, the Lorimer profile is always preferred over the
Gaussian disk. 
Moreover, the inferred parameters agree within errors between different datasets, but 
get more tightly constrained by larger datasets (see Figs~\ref{fig:bench},
\ref{fig:cp_2PC} and \ref{fig:cp_ALL}).

This leads us to the somewhat surprising conclusion
that for the purpose of our analysis there is no strong bias when including only $\gamma$-ray detected MSPs in the
analysis. 
A priori this is not obvious, since all but one source have radio counterparts which could lead
to a selection bias which cannot be efficiently accounted for in the detection sensitivity.
Moreover, for all but one of the unassociated sources we do not have distance priors. 
This analysis however shows that we can derive consistent constraints whether or not
distance information is included (also see \cite{Hooper:2015jlu}).

In the future, it would be interesting to include a larger sample of likely
pulsar candidates in order to constrain the luminosity function down to lower fluxes.
In particular, without radio counterpart, it is difficult to confirm $\gamma$-ray
pulsations in blind searches \citep{Clark:2018zsp}.
One possibility would be an update of the work by \cite{Parkinson:2016oab} using a
larger source catalog.
In addition, \cite{Fermi-LAT:2017yoi}\footnote{Also see \cite{Bartels:2017xba}.} propose a potentially powerful technique which
classifies unassociated sources as likely pulsar candidates and which uses 
a customized detection efficiency.

\subsection{Implications for the Galactic Center Excess}
We tested for the presence of a bulge MSP population by including an additional component in our analysis (Sect.~\ref{sec:methods}), but find no evidence for the presence of such a population (Sect.~\ref{sec:results}).
This analysis assumes that bulge MSPs follow the same luminosity function
as disk MSPs. 
Using the same assumption and the observed GCE intensity we can also
estimate how many MSPs from the bulge should have been detected.
We use a GCE intensity of $2.3\times 10^{-9}\mathrm{\,erg\,cm^{-2}\,s^{-1}}$
\citep{Bartels:2017vsx} and distance to the GCE of $R_\odot=8.5\mathrm{\,kpc}$
to normalize the bulge population. Using the best-fit detection efficiency 
and luminosity function of our benchmark model
we estimate that $4.5$ sources should have been detected. 
Within the 95\% containment interval of the full posterior 
the number of bulge MSP detections ranges 
from being fewer than one to more than a dozen.
Using the
dataset that includes unassociated sources, this number goes up to $5.5$.
Similar numbers are obtained for the log-normal luminosity function.
We therefore agree with \cite{Ploeg:2017vai} that the MSP
interpretation of the GCE is
consistent with the luminosity function derived from MSPs in the Galactic disk. 

We find that opposite conclusions are
driven by the high-luminosity tail of the luminosity function. 
At the distance of the GC 
mostly sources with luminosities
$\gtrsim 2\times 10^{34}\mathrm{\,erg\,s^{-1}}$ can be detected. 
\cite{Hooper:2015jlu} find relatively more bright sources ($\geq 10^{34}\mathrm{\,erg\,s^{-1}}$), and
thus a higher number of expected bulge detections, compared to 
this work and \cite{Ploeg:2017vai}.
Similarly, the MSP population in globular clusters has about an order-of-magnitude 
higher mean luminosity than what we derive for the disk
\citep{Hooper:2016rap}.
The treatment of the flux threshold only has a mild impact.
Here and in \cite{Ploeg:2017vai} both $K_\mathrm{th}$
and $\sigma_\mathrm{th}$ are left free in the fit. However,
\cite{Hooper:2015jlu} fix $\sigma_\mathrm{th}=0.9$, which is 
larger than our best-fit value. Although this leads
to a larger acceptance of dim sources,
the detection probabilities at $\gtrsim 3\times 10^{34}\mathrm{\,erg\,s^{-1}}$ 
are very similar.

It should be mentioned that all but one of the $\gamma$-ray detected MSPs
have radio counterparts. It is notoriously difficult to detect MSPs in radio
near the Galactic Center due to the large scatter 
broadening of pulsed emission \citep[e.g.~][]{Calore:2015bsx}.
Since we apply a detection threshold based on $\gamma$-ray flux this does not directly
take into account the decreasing sensitivity of radio searches with increasing distance.
Consequently, if bulge MSPs are present in our full sample it is not unlikely that they are
all unassociated sources. In the near future, the radio sensitivity for searches of bulge MSPs
should increase significantly, allowing for the detection of this component in 
radio \citep{Calore:2015bsx}.

If the GCE originates from MSPs in the disk 
we find a bulge-to-disk ($B/D$) luminosity (flux) ratio of $B/D\sim 1.3\,(0.5$).
The ratio of luminosity-to-stellar mass in the bulge is 
$2.2\times 10^{27}\mathrm{\,erg\,s^{-1}\,M_\odot^{-1}}$ compared to
$2.9\times10^{26}\mathrm{\,erg\,s^{-1}\,M_\odot^{-1}}$ in the disk.
Therefore, the bulge appears to host approximately eight times
more MSPs per unit stellar mass than the disk, consistent with the results 
from \cite{Bartels:2017vsx}.

\subsection{Completeness}
\label{sec:completeness}
We discuss the completeness we obtain from our analysis, i.e.~the number of detected
sources over the total number of sources in the disk,
and compare it to the results of \cite{Winter:2016wmy}.
Although \cite{Winter:2016wmy} find a comparable parameterization of the luminosity function,
their normalization and therefore total luminosity is about a factor $7$ larger 
than what we find \citep{Winter:2016wmy, Eckner:2017oul}. This difference can be ascribed to the fact that our analysis
yields a larger completeness by about a factor $\sim 10$ 
at the peak of the luminosity function 
($L\sim10^{33}\mathrm{\,erg\,s^{-1}}$). It should be taken into account that we use
a larger sample of MSPs, 96 versus 66 in \cite{Winter:2016wmy}.
Naively rescaling by this ratio	still 
leaves a factor $\sim 7$ higher completeness.

\begin{figure}[h]
  \centering
  \includegraphics[width=\linewidth]{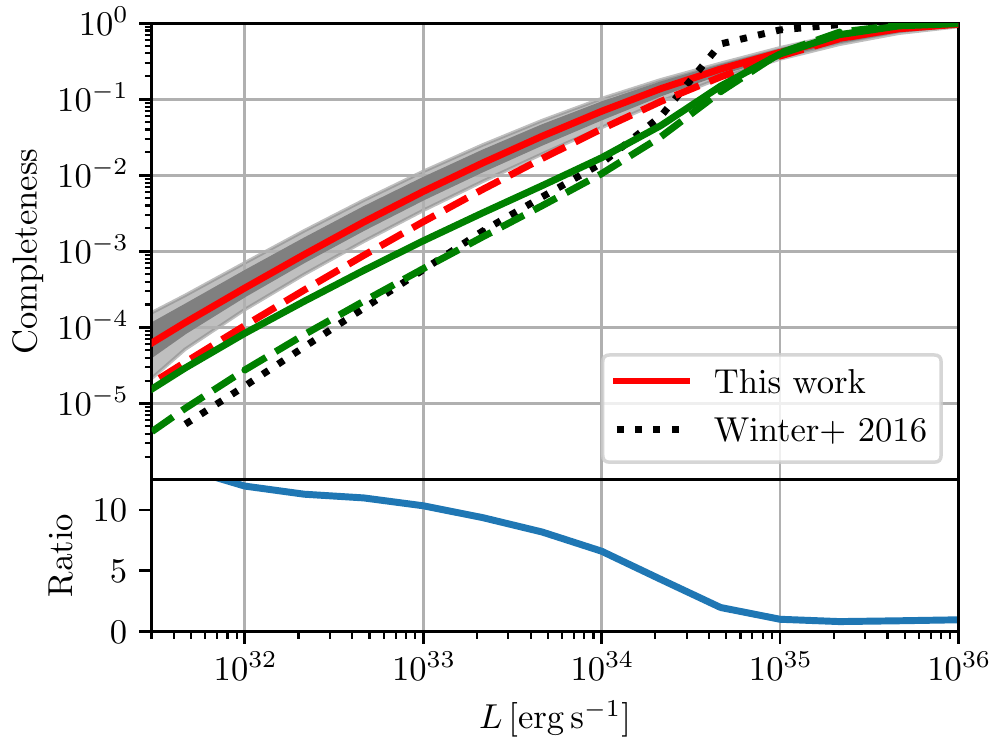}
  \caption{\textit{Top panel:} Comparison of completeness between this work (solid red line)
  and \cite{Winter:2016wmy} (black dotted line). 
  The dark (light) grey band shows the 68\% (95\%) containment 
  interval of the completeness obtained in this work.
  Color and line style indicate spatial distribution and flux threshold respectively. 
  Red (green) use the spatial distribution 
  from this work (\cite{Winter:2016wmy}). The solid (dashed) lines use the flux threshold
  from this work (\cite{Winter:2016wmy}).
  The green dashed line shows our reproduction of the MC simulation of \cite{Winter:2016wmy},
  using their spatial distribution and flux sensitivity.
  \textit{Bottom panel:} ratio of the red-solid over the green-dashed line.
  }
  \label{fig:completeness}
\end{figure}

We find the reason for the difference in completeness to be twofold.
First, \cite{Winter:2016wmy} estimate completeness by performing a Monte-Carlo (MC) simulation.
They randomly draw pulsars at a given luminosity and assign it a position
by drawing from a Lorimer profile with $B=0$, $C = 2.8$ and $z_s = 0.6$
\citep{Story:2007xy, Gregoire:2013yta,Winter:2016wmy}. 
In fact, $C$ and $z_s$ 
are themselves also drawn from log-normal distribution. This profile is consistent 
with our best-fit value at $\sim 2\sigma$. 
We compare the impact this has on the completeness by running a MC simulation
drawing sources at different luminosities and assigning them 
spatial positions based on the distribution assumed in \cite{Winter:2016wmy}
and our benchmark distribution. 
With the spatial profile from \cite{Winter:2016wmy}
the MSPs are on average slightly further away compared to our benchmark spatial profile.
Consequently, the flux received from each source is about a factor $\sim 2$ dimmer,
which so happens to also result in a loss of completeness by a factor $\sim 2$.
This is displayed Fig.~\ref{fig:completeness} as the difference between
the green and red lines with the same linestyle.
Second, the detection threshold applied by \cite{Winter:2016wmy} is based on
latitude dependent flux threshold in Fig.~17 from \cite{TheFermi-LAT:2013ssa}, 
whereas we use the map in Fig.~16 of that same work.
In our MC simulation we also compare these two detection sensitivities.
In Fig.~\ref{fig:completeness} this is shown by the difference 
between the solid (our detection sensitivity) and dashed 
(sensitivity threshold from \cite{Winter:2016wmy}) lines of the same color.
We find that our sensitivity function yields a larger completeness.
Finally, we note that in our MC simulation the red solid line corresponds
to our benchmark model and the green solid line to our reproduction 
of the completeness from \cite{Winter:2016wmy}, with their spatial distribution 
and flux threshold. In the bottom panel we show the ratio of the
red-solid line (our work) over the green-dashed line 
\citep[our MC reproduction of][which agrees very well]{Winter:2016wmy}.

A merit of our analysis is that it is fully self-consistent in that we model
the spatial-distribution, luminosity function and flux sensitivity simultaneously.
Therefore, we consider the grey band in Fig.~\ref{fig:completeness}
to be the
most trustworthy representation of the completeness. 
It shows the 68\% and 95\% containment interval of the completeness for our benchmark model.
We construct it
by sampling spatial and flux-sensitivity parameters from the full \texttt{MultiNest} posterior and consecutively running a MC simulation to estimate the completeness for
each point.

The estimated completeness at the peak of the
luminosity function has a large impact on the derived ratio of emission per unit
stellar mass and therefore the expected unresolved flux. 
Any conclusions that relies directly on the completeness by 
estimating the luminosity of MSPs in a given environment by 
applying the luminosity-per-stellar mass from the disk
is affected by this uncertainty. 
With our estimate of the completeness we expect MSPs from the disk
to contribute $\mathcal{O}(1\%)$ to the total $\gamma$-ray flux from
$1$--$10\mathrm{\,GeV}$, where the MSP spectrum is most pronounced \citep{Mccann:2014dea}.
Moreover, 
studies of dwarf spheroidal galaxies \citep{Winter:2016wmy}, 
Andromeda \citep{Eckner:2017oul} and the Galactic Bulge 
\citep{Macias:2016nev,Bartels:2017vsx} are also affected by our estimated completeness\footnote{The uncertainty in the completeness and its impact were already briefly discussed in \citep{Bartels:2017vsx, Eckner:2017oul}}.

\section{Conclusion}
\label{sec:conclusion}
We have performed 
a Bayesian-unbinned likelihood analysis and -- for the first time in this 
context -- done Bayesian model comparison in order to constrain
the properties of the Galactic population of $\gamma$-ray MSPs, self-consistently taking into account various sources of uncertainties.
We used a sample of $96$ $\gamma$-ray detected MSPs, but verified that our results 
remain similar under the inclusion of an additional $69$ well-motivated MSP candidates.
In order to deal with distance uncertainties 
we developed a novel method to construct PDFs for the distance to individual pulsars and the distance proxies.
We use the YMW16 electron-density model to construct a PDF for the dispersion
measure given the true distance to a pulsar by
sampling from the model its $35$ free parameters. We therefore take into
account the uncertainties on the derived parameters within the electron density model
\citep{2017ApJ...835...29Y}.
Distance and flux uncertainties were then marginalized over.
The normalization and variance of the flux-detection threshold were treated as free
parameters in our analysis. Results for different parameterizations
of the luminosity function and spatial profile are compared by computing Bayes factors.

We find that a Lorimer-disk profile is preferred over a disk with a Gaussian radial profile,
although the parameters are only loosely constrained. There is clear evidence for a turnover
in the luminosity function, ruling out a single power-law parameterization (with hard or super-exponential cutoff).
Instead, both a broken-power-law and a log-normal function provide good fits to the luminosity function.

Our analysis suggests the presence of $\sim 2\times 10^{4}$ MSPs in the Galactic disk. However, within uncertainties this
number could be as large as $\sim 10^{5}$.
These numbers are in agreement with the expected MSP population derived using 
radio catalogs \citep{Cordes:1997my, 1998MNRAS.295..743L, Levin:2013usa}.

Contrary to previous claims \citep{Hooper:2013nhl,Cholis:2014lta,Hooper:2015jlu},
we find the MSP interpretation of the GCE to be fully compatible with the characteristics of the disk MSPs.
Therefore, we agree with the findings of \cite{Yuan:2014rca,Petrovic:2014xra,Ploeg:2017vai}. 
Our characterization of the luminosity function and detection sensitivity suggest that if the luminosity function of the bulge MSP population is identical to that of the disk MPS, and if 100\% of the GCE is due to MSPs,  
only a handful of sources should have been detected from the bulge, whereas
in the past larger numbers were suggested.
We explicitly tested for the presence of a bulge component in our analysis, but find that we
currently lack sensitivity to place interesting constraints on the bulge population of MSPs.
In the future, an extension of the work by \cite{Parkinson:2016oab} or a dedicated analysis to characterize unassociated sources as likely pulsars
\citep{Fermi-LAT:2017yoi} can be potentially powerful methods to constrain the bulge population.

At the peak of the luminosity function we find a higher detection completeness than previous
work \citep{Winter:2016wmy}. Consequently, our luminosity-per-stellar-mass ratio of
$\sim 3\times 10^{26}\mathrm{\,erg\,s^{-1}\,M_\odot^{-1}}$ is significantly smaller than what has been derived
in other works \citep{Winter:2016wmy, Macias:2016nev, Eckner:2017oul}. It should be mentioned that the completeness suffers from considerable uncertainties due to its dependence on the detection sensitivity,
spatial profile and luminosity function.

The results presented in this work have direct implications for the detectability of a diffuse disk MSP component due to unresolved sources, their contribution
to the isotropic $\gamma$-ray background \citep{FaucherGiguere:2009df,SiegalGaskins:2010mp,Calore:2014oga}, the bulge-to-disk ratio
of MSPs \citep{Macias:2016nev,Bartels:2017vsx}, the expected emission
from dwarf galaxies \citep{Winter:2016wmy}, and the detectability of MSPs in external Galaxies
such as M31 \citep{Eckner:2017oul}. 

Although the properties of the galactic disk MSP population are the main topic of this paper, the methods we describe can be applied directly to any population of astrophysical sources where unassociated sources are present and distances uncertainties are large, a situation commonly found in population analyses.

\acknowledgements
We thank Jason Hessels, Sebastian Liem and Dick Manchester for discussion,
Dan Hooper, Pasquale Serpico and Gabrijela Zaharijas for feedback on the manuscript,
and David Smith for providing the sensitivity map.
This work was carried out on the Dutch national e-infrastructure with the support of SURF Cooperative.
This research is funded by NWO through the VIDI research program
"Probing the Genesis of Dark Matter" (680-47-532; TE, CW) 
and through a GRAPPA-PhD fellowship (022.004.017; RB).

\bibliographystyle{aasjournal_RTB}
\bibliography{msp}

\appendix
\section{Millisecond pulsar sample}
In Table \ref{tab:sources} we show the source list used in this work.
The full list is available at \url{http://github.com/tedwards2412/MSPDist}.
We separated the table in $\gamma$-ray detected pulsars and unassociated sources.
For each source we give the position, $\gamma$-ray flux ($0.1-100\mathrm{\,GeV}$),
dispersion measure and/or parallax if available. 
Finally, the catalogs in which the sources appear are given.
\startlongtable
\begin{deluxetable}{l|cccccc}

\tablecaption{\label{tab:sources}Millisecond pulsar sample separated into $\gamma$-ray detected MSPs
and MSP candidates from \cite{Parkinson:2016oab} (see text for details).
The different columns provide respectively: the name of the source,
Galactic longitude and latitude in degrees, $\gamma$-ray flux in the 
range $0.1-100\mathrm{\,GeV}$, the dispersion measure and/or parallax from that ATNF
\citep{Manchester:2004bp} if available,
and finally a reference to the relevant catalogs.}
\tablehead{
\colhead{Name} & \colhead{$\ell$} & \colhead{$b$} &
\colhead{Flux}
& \colhead{DM} & \colhead{Parallax}
& \colhead{Catalogs\tablenotemark{a}}
\\
\colhead{} & \colhead{$\left[\mathrm{deg}\right]$} & \colhead{$\left[\mathrm{deg}\right]$} &
\colhead{$\left[10^{-12}\mathrm{\,erg\,cm^{-2}\,s^{-1}}\right]$} 
& \colhead{$\left[\mathrm{cm^{-3}\,pc}\right]$}
& \colhead{$\left[\mathrm{mas}\right]$}
}
\startdata
\hline
\multicolumn{7}{c}{$\gamma$-ray pulsars (96)}\\
\hline
          J0023+0923 & 111.5 & -52.9 & $7.28\pm0.81$ & 14.33 & $0.93\pm0.16 $ & 1,2,3,4,5\\
          J0030+0451 & 113.1 & -57.6 & $60.68\pm1.51$ & 4.34 & $3.08\pm0.09 $ & 1,2,3,4,5\\
          J0034-0534 & 111.5 & -68.1 & $18.04\pm1.02$ & 13.77 & * & 1,2,3,4,5\\
          J0101-6422 & 301.2 & -52.7 & $12.45\pm0.85$ & 11.93 & * & 1,2,3,4,5\\
          J0102+4839 & 124.9 & -14.2 & $16.76\pm1.39$ & 53.50 & * & 1,2,3,5\\
          J0218+4232 & 139.5 & -17.5 & $48.14\pm1.80$ & 61.25 & $0.16\pm0.09 $ & 1,2,3,4,5\\
          J0248+4230\tablenotemark{f} & 144.9 & -15.3 & $5.21\pm0.81$ & 48.2 & * & 2,4,5\\
            J0251+26 & 153.9 & -29.5 & $6.87\pm0.99$ & 20.00 & * & 2,3,4,5\\
            J0308+74\tablenotemark{b} & 131.7 & 14.2 & $14.57\pm0.79$ & 6.35 & * & 2,3,5\\
          J0318+0253 & 178.4 & -43.6 & $5.71\pm0.74$ & 26. & * & 2,3,4,5\\
          J0340+4130 & 153.8 & -11.0 & $22.24\pm1.33$ & 49.59 & $0.7\pm0.5 $ & 1,2,3,4,5\\
          J0437-4715 & 253.4 & -42.0 & $17.87\pm0.88$ & 2.64 & $6.37\pm0.09 $ & 1,2,3,4,5\\
            J0533+67 & 144.8 & 18.2 & $9.57\pm0.89$ & 57.40 & * & 2,3,5\\
            J0605+37 & 174.2 & 8.0 & $6.88\pm0.95$ & 21.00 & * & 2,3,5\\
          J0610-2100 & 227.7 & -18.2 & $11.48\pm1.07$ & 60.67 & * & 1,2,3,4,5\\
          J0613-0200 & 210.4 & -9.3 & $33.57\pm1.64$ & 38.78 & $0.93\pm0.2 $ & 1,2,3,4,5\\
          J0614-3329 & 240.5 & -21.8 & $110.80\pm2.36$ & 37.05 & * & 1,2,3,4,5\\
            J0621+25 & 187.1 & 5.1 & $11.04\pm1.50$ & 83.60 & * & 2,3,4\\
         J0737-3039A\tablenotemark{c} & 245.2 & -4.5 & $4.00\pm1.00$ & 48.92 & * & 5\\
          J0740+6620 & 149.7 & 29.6 & $4.77\pm0.68$ & 14.96 & $2.3\pm0.7 $ & 2,4,5\\
          J0751+1807 & 202.8 & 21.1 & $13.04\pm0.97$ & 30.25 & $0.82\pm0.17 $ & 1,2,3,4,5\\
          J0931-1902 & 251.0 & 23.0 & $3.00\pm0.86$ & 41.49 & $1.2\pm0.9 $ & 2,4,5\\
            J0955-61 & 283.7 & -5.7 & $8.24\pm1.27$ & 160.70 & * & 2,5\\
          J1012-4235 & 274.2 & 11.2 & $7.48\pm1.12$ & 71.60 & * & 2,3,4,5\\
          J1023+0038 & 243.4 & 45.8 & $5.35\pm0.97$ & 14.32 & $0.731\pm0.022 $ & 3,4\\
          J1024-0719 & 251.7 & 40.5 & $3.58\pm0.52$ & 6.48 & $0.8\pm0.3 $ & 1,2,3,4,5\\
          J1035-6720\tablenotemark{d} & 290.4 & -7.8 & $25.94\pm1.47$ & 84.16 & * & 2,3,4,5\\
          J1036-8317 & 298.9 & -21.5 & $5.78\pm0.95$ & 27.00 & * & 2,4,5\\
          J1124-3653 & 284.1 & 22.8 & $13.16\pm1.09$ & 44.90 & * & 1,2,3,5\\
          J1125-5825 & 291.8 & 2.6 & $14.51\pm2.66$ & 124.79 & * & 1,2,3,4,5\\
          J1137+7528\tablenotemark{e} & 129.1 & 40.8 & $2.28\pm0.59$ & 29.1702 & * & 2,4,5\\
          J1142+0119 & 267.6 & 59.4 & $6.24\pm0.82$ & 19.20 & * & 2,3,5\\
          J1207-5050 & 295.9 & 11.4 & $7.89\pm1.16$ & 50.60 & * & 2,3,4,5\\
          J1227-4853 & 299.0 & 13.8 & $41.36\pm1.67$ & 43.42 & * & 2,3,4,5\\
          J1231-1411 & 295.5 & 48.4 & $102.86\pm2.12$ & 8.09 & * & 1,2,3,4,5\\
          J1301+0833 & 310.8 & 71.3 & $10.63\pm0.97$ & 13.20 & * & 2,3,4,5\\
            J1302-32 & 305.6 & 29.8 & $11.30\pm1.13$ & 26.20 & * & 2,3,5\\
          J1311-3430 & 307.7 & 28.2 & $64.69\pm1.89$ & 37.84 & * & 2,3,4,5\\
          J1312+0051 & 314.9 & 63.2 & $16.50\pm1.10$ & 15.30 & * & 2,3,5\\
          J1431-4715 & 320.1 & 12.3 & $6.41\pm0.95$ & 59.35 & * & 4,5\\
          J1446-4701 & 322.5 & 11.4 & $12.55\pm1.30$ & 55.83 & * & 1,2,3,4,5\\
          J1455-3330 & 330.8 & 22.5 & $2.15\pm0.50$ & 13.57 & $0.99\pm0.22 $ & 4,5\\
          J1513-2550\tablenotemark{e} & 338.8 & 27.0 & $7.03\pm0.98$ & 46.86 & * & 2,3,4,5\\
          J1514-4946 & 325.2 & 6.8 & $42.81\pm2.12$ & 31.05 & * & 1,2,3,4,5\\
          J1536-4948 & 328.2 & 4.8 & $87.43\pm3.05$ & 38.00 & * & 2,3,5\\
          J1543-5149 & 327.9 & 2.7 & $21.83\pm2.60$ & 50.93 & * & 2,4,5\\
          J1544+4937 & 79.2 & 50.2 & $3.58\pm0.64$ & 23.23 & * & 2,4,5\\
          J1552+5437 & 85.6 & 47.2 & $4.53\pm0.64$ & 22.90 & * & 2,4,5\\
          J1600-3053 & 344.1 & 16.5 & $6.16\pm1.07$ & 52.33 & $0.5\pm0.08 $ & 1,2,3,4,5\\
          J1614-2230 & 352.6 & 20.2 & $23.37\pm1.49$ & 34.92 & $1.5\pm0.1 $ & 1,2,3,4,5\\
          J1622-0315\tablenotemark{e} & 10.8 & 30.7 & $10.15\pm1.28$ & 21.4 & * & 2,3,4,5\\
          J1628-3205 & 347.4 & 11.5 & $12.12\pm1.48$ & 42.10 & * & 2,3,4,5\\
            J1630+37 & 60.2 & 43.3 & $6.86\pm1.01$ & 14.10 & * & 2,3,5\\
          J1640+2224 & 41.0 & 38.3 & $2.59\pm0.45$ & 18.46 & $0.66\pm0.07 $ & 4,5\\
          J1658-5324 & 334.9 & -6.6 & $20.32\pm1.99$ & 30.81 & * & 1,2,3,4,5\\
          J1713+0747 & 28.8 & 25.2 & $9.41\pm1.25$ & 15.92 & $0.81\pm0.03 $ & 1,2,3,4,5\\
          J1730-2304 & 3.2 & 6.0 & $12.97\pm2.38$ & 9.62 & $1.19\pm0.27 $ & 4,5\\
          J1732-5049 & 340.0 & -9.4 & $8.52\pm1.34$ & 56.84 & * & 2,4,5\\
          J1741+1351 & 37.9 & 21.6 & $5.68\pm1.06$ & 24.20 & $0.56\pm0.13 $ & 1,2,3,4,5\\
          J1744-1134 & 14.8 & 9.2 & $39.16\pm2.18$ & 3.14 & $2.53\pm0.07 $ & 1,2,3,4,5\\
          J1744-7619 & 317.1 & -22.5 & $22.50\pm1.31$ & * & * & 2,3,4,5\\
          J1745+1017 & 34.9 & 19.3 & $10.56\pm1.48$ & 23.97 & * & 2,3,4,5\\
          J1747-4036 & 350.2 & -6.4 & $15.97\pm1.79$ & 152.96 & * & 1,2,3,4,5\\
            J1805+06 & 33.4 & 13.0 & $5.51\pm0.99$ & 65.00 & * & 2,3,4,5\\
          J1810+1744 & 44.6 & 16.8 & $22.38\pm1.37$ & 39.70 & * & 1,2,3,4,5\\
          J1811-2405 & 6.9 & -2.5 & $21.79\pm4.30$ & 60.60 & * & 2,4\\
          J1816+4510 & 72.9 & 24.8 & $12.13\pm0.93$ & 38.89 & * & 2,3,4,5\\
          J1832-0836 & 23.0 & 0.2 & $15.27\pm2.99$ & 28.19 & * & 4,5\\
          J1843-1113 & 22.0 & -3.4 & $19.81\pm2.80$ & 59.96 & $0.69\pm0.33 $ & 2,4,5\\
          J1855-1436\tablenotemark{e} & 20.4 & -7.6 & $7.85\pm1.00$ & 109.2 & * & 4,5\\
          J1858-2216 & 13.6 & -11.4 & $8.33\pm1.09$ & 26.60 & * & 1,2,3,5\\
          J1902-5105 & 345.6 & -22.4 & $21.47\pm1.16$ & 36.25 & * & 1,2,3,4,5\\
            J1902-70 & 324.4 & -26.5 & $12.28\pm0.99$ & 19.50 & * & 2,3,5\\
            J1909+21 & 53.7 & 5.8 & $7.01\pm1.10$ & 62.00 & * & 4,5\\
          J1921+0137\tablenotemark{e} & 37.8 & -5.9 & $15.92\pm1.92$ & 104.9 & * & 2,3,4,5\\
          J1939+2134 & 57.5 & -0.3 & $9.18\pm3.32$ & 71.02 & $0.22\pm0.08 $ & 1,4,5\\
          J1946-5403 & 343.9 & -29.6 & $11.29\pm0.92$ & 23.70 & * & 2,3,4,5\\
          J1959+2048 & 59.2 & -4.7 & $17.91\pm1.54$ & 29.12 & * & 1,2,3,4,5\\
          J2017+0603 & 48.6 & -16.0 & $34.97\pm1.69$ & 23.92 & $0.4\pm0.3 $ & 1,2,3,4,5\\
          J2017-1614\tablenotemark{e} & 27.3 & -26.2 & $10.40\pm1.20$ & 25.4380 & * & 2,3,4,5\\
          J2042+0246\tablenotemark{e} & 49.0 & -23.0 & $3.61\pm0.55$ & 9.2694 & * & 2,4,5\\
          J2043+1711 & 61.9 & -15.3 & $30.22\pm1.41$ & 20.76 & $0.64\pm0.08 $ & 1,2,3,4,5\\
          J2047+1053 & 57.1 & -19.6 & $3.56\pm0.58$ & 34.60 & * & 1,2,3,5\\
          J2051-0827 & 39.2 & -30.5 & $3.18\pm0.52$ & 20.73 & * & 1,2,4,5\\
          J2052+1218 & 59.1 & -20.0 & $6.53\pm1.04$ & 42.00 & * & 2,4,5\\
          J2124-3358 & 10.9 & -45.4 & $39.40\pm1.39$ & 4.60 & $2.4\pm0.4 $ & 1,2,3,4,5\\
          J2129-0429 & 48.9 & -36.9 & $10.50\pm1.06$ & 16.90 & * & 2,3,4,5\\
          J2205+6015\tablenotemark{f} & 103.7 & 3.8 & $7.50\pm1.52$ & 157.6 & * & 4,5\\
          J2214+3000 & 86.9 & -21.7 & $33.00\pm1.24$ & 22.55 & $2.3\pm0.7 $ & 1,2,3,4,5\\
          J2215+5135 & 99.9 & -4.2 & $13.75\pm1.14$ & 69.20 & * & 1,2,3,4,5\\
          J2234+0944 & 76.3 & -40.4 & $8.28\pm1.01$ & 17.8 & $1.3\pm0.5 $ & 2,3,4,5\\
          J2241-5236 & 337.4 & -54.9 & $30.97\pm1.22$ & 11.41 & * & 1,2,3,4,5\\
          J2256-1024 & 59.2 & -58.2 & $7.66\pm0.78$ & 13.80 & * & 2,3,4,5\\
          J2302+4442 & 103.4 & -14.0 & $38.10\pm1.40$ & 13.73 & * & 1,2,3,4,5\\
          J2310-0555\tablenotemark{e} & 69.7 & -57.9 & $3.48\pm0.56$ & 15.5139 & * & 2,4,5\\
          J2339-0533\tablenotemark{e} & 81.3 & -62.5 & $30.06\pm1.39$ & 8.72 & * & 2,3,4,5\\
\hline
\multicolumn{7}{c}{other sources (69)}\\
\hline
        J0039.3+6256 & 121.6 & 0.1 & $9.11\pm1.14$ &          * & * & 2,3,4\\
        J0212.1+5320 & 134.9 & -7.6 & $17.14\pm1.56$ &          * & * & 2,3,4\\
        J0238.0+5237 & 138.8 & -6.9 & $11.60\pm1.21$ &          * & * & 2,3,4\\
        J0312.1-0921 & 191.5 & -52.4 & $5.23\pm0.84$ &          * & * & 2,3,4\\
        J0336.1+7500 & 133.1 & 15.5 & $9.97\pm1.04$ &          * & * & 2,3,4\\
        J0401.4+2109 & 171.4 & -23.3 & $6.27\pm1.09$ &          * & * & 3,4\\
        J0523.3-2528 & 228.2 & -29.8 & $19.91\pm1.24$ &          * & * & 2,3,4\\
       J0542.5-0907c & 213.4 & -19.4 & $13.64\pm1.81$ &          * & * & 3,4\\
        J0545.6+6019 & 152.5 & 15.7 & $7.87\pm0.95$ &          * & * & 2,3,4\\
        J0737.2-3233 & 246.8 & -5.5 & $13.83\pm1.52$ &          * & * & 2,3,4\\
        J0744.1-2523 & 241.3 & -0.7 & $23.86\pm1.78$ &          * & * & 2,3,4\\
        J0744.8-4028 & 254.6 & -8.0 & $9.40\pm1.36$ &          * & * & 3,4\\
        J0758.6-1451 & 234.0 & 7.6 & $7.30\pm1.06$ &          * & * & 2,3,4\\
        J0802.3-5610 & 269.9 & -13.2 & $13.01\pm1.18$ &          * & * & 2,3,4\\
        J0826.3-5056 & 267.4 & -7.4 & $10.66\pm1.59$ &          * & * & 3,4\\
        J0838.8-2829 & 250.6 & 7.8 & $12.74\pm1.20$ &          * & * & 2,3,4\\
        J0933.9-6232 & 282.2 & -7.9 & $12.27\pm1.06$ &          * & * & 2,3,4\\
        J0953.7-1510 & 251.9 & 29.6 & $5.85\pm0.71$ &          * & * & 2,3,4\\
        J0954.8-3948 & 269.8 & 11.5 & $18.29\pm1.23$ &          * & * & 2,3,4\\
        J0957.6+5523 & 158.6 & 47.9 & $95.86\pm2.73$ &          * & * & 2,3,4\\
        J1119.9-2204 & 276.5 & 36.1 & $16.85\pm1.03$ &          * & * & 2,3,4\\
        J1136.1-7411 & 297.8 & -12.1 & $11.18\pm1.18$ &          * & * & 2,3,4\\
        J1207.6-4537 & 295.0 & 16.6 & $4.17\pm0.93$ &          * & * & 3,4\\
        J1208.0-6901 & 299.0 & -6.5 & $7.50\pm1.25$ &          * & * & 3,4\\
        J1225.9+2953 & 185.2 & 83.8 & $8.70\pm0.97$ &          * & * & 2,3\\
        J1306.4-6043 & 304.8 & 2.1 & $35.12\pm2.50$ &          * & * & 2,3,4\\
        J1325.2-5411 & 307.9 & 8.4 & $10.75\pm1.65$ &          * & * & 2,3,4\\
        J1329.8-6109 & 307.6 & 1.4 & $16.47\pm2.39$ &          * & * & 2,3,4\\
        J1400.2-2413 & 322.4 & 36.0 & $5.82\pm0.99$ &          * & * & 2,3,4\\
        J1400.5-1437 & 326.9 & 45.0 & $9.36\pm1.09$ & 4.93 & $3.6\pm1.1 $ & 2,3,4\\
        J1412.3-6635 & 310.9 & -5.0 & $8.21\pm1.46$ &          * & * & 3,4\\
        J1458.7-2120 & 338.6 & 32.6 & $7.05\pm1.05$ &          * & * & 2,3,4\\
        J1539.2-3324 & 338.8 & 17.5 & $11.56\pm1.03$ &          * & * & 2,3,4\\
        J1544.6-1125 & 356.2 & 33.0 & $13.54\pm1.40$ &          * & * & 2,3,4\\
        J1600.3-5810 & 325.8 & -3.9 & $5.50\pm1.40$ &          * & * & 3,4\\
        J1616.8-5343 & 330.5 & -2.2 & $26.48\pm2.62$ &          * & * & 2,3,4\\
        J1624.2-3957 & 341.1 & 6.6 & $13.09\pm2.53$ &          * & * & 3,4\\
        J1625.1-0021 & 13.9 & 31.8 & $18.38\pm1.26$ &          * & * & 2,3,4\\
        J1630.2-1052 & 4.9 & 24.8 & $6.71\pm1.32$ &          * & * & 2,3,4\\
        J1641.5-5319 & 333.3 & -4.6 & $18.42\pm2.24$ &          * & * & 2,3,4\\
        J1653.6-0158 & 16.6 & 24.9 & $33.71\pm1.83$ &          * & * & 2,3,4\\
        J1702.8-5656 & 332.4 & -9.2 & $32.04\pm1.66$ &          * & * & 2,3,4\\
        J1717.6-5802 & 332.6 & -11.5 & $12.39\pm1.30$ &          * & * & 2,3,4\\
        J1722.7-0415 & 18.5 & 17.5 & $12.33\pm2.08$ &          * & * & 2,3,4\\
        J1730.6-0357 & 19.8 & 16.0 & $6.44\pm1.25$ &          * & * & 2,3,4\\
        J1740.5-2642 & 1.3 & 2.1 & $16.77\pm2.51$ &          * & * & 3,4\\
        J1743.9-1310 & 13.3 & 8.5 & $8.23\pm1.79$ &          * & * & 3,4\\
        J1748.5-3912 & 351.5 & -5.8 & $16.26\pm1.94$ &          * & * & 3,4\\
        J1749.7-0305 & 23.0 & 12.2 & $12.84\pm1.88$ &          * & * & 3,4\\
        J1753.6-4447 & 347.1 & -9.4 & $9.36\pm1.28$ &          * & * & 2,3,4\\
        J1759.2-3848 & 352.9 & -7.4 & $8.92\pm1.69$ &          * & * & 2,3,4\\
        J1808.3-3357 & 358.1 & -6.7 & $8.72\pm1.44$ &          * & * & 2,3,4\\
        J1823.2-4722 & 347.1 & -15.2 & $4.82\pm1.01$ &          * & * & 3,4\\
        J1827.7+1141 & 40.8 & 10.5 & $7.57\pm1.31$ &          * & * & 2,3,4\\
        J1830.8-3136 & 2.4 & -9.8 & $6.77\pm1.35$ &          * & * & 2,3,4\\
        J1908.8-0130 & 33.6 & -4.6 & $7.12\pm0.94$ &          * & * & 2,3,4\\
        J1918.2-4110 & 356.8 & -22.2 & $21.61\pm1.86$ &          * & * & 2,3,4\\
        J1950.2+1215 & 50.7 & -7.1 & $16.13\pm1.75$ &          * & * & 2,3,4\\
        J2004.4+3338 & 70.7 & 1.2 & $43.07\pm2.79$ &          * & * & 2,3,4\\
        J2006.6+0150 & 43.4 & -15.8 & $4.17\pm1.02$ &          * & * & 3,4\\
        J2026.8+2813 & 68.8 & -5.8 & $7.66\pm1.50$ &          * & * & 3,4\\
        J2035.0+3634 & 76.6 & -2.3 & $12.32\pm1.82$ &          * & * & 2,3,4\\
        J2039.6-5618 & 341.2 & -37.2 & $17.11\pm1.38$ &          * & * & 2,3,4\\
        J2043.8-4801 & 351.7 & -38.3 & $7.35\pm0.93$ &          * & * & 2,3,4\\
        J2112.5-3044 & 14.9 & -42.4 & $19.01\pm1.39$ &          * & * & 2,3,4\\
        J2117.6+3725 & 82.8 & -8.3 & $12.76\pm1.31$ &          * & * & 2,3,4\\
        J2133.0-6433 & 328.7 & -41.3 & $3.97\pm0.67$ &          * & * & 2,3,4\\
        J2212.5+0703 & 68.7 & -38.6 & $9.03\pm1.03$ &          * & * & 2,3,4\\
        J2250.6+3308 & 95.7 & -23.3 & $5.27\pm0.87$ &          * & * & 3,4\\
\enddata
\tablenotetext{a}{(1) in 2PC \citep{TheFermi-LAT:2013ssa}; 
(2) in 2FGL \citep{Fermi-LAT:2011yjw};
(3) in 3FGL \citep{Acero:2015hja};
(4) in FL8Y \url{https://fermi.gsfc.nasa.gov/ssc/data/access/lat/fl8y/};
(5) in the Public list of Fermi-LAT detected $\gamma$--ray
pulsars \url{https://confluence.slac.stanford.edu/display/GLAMCOG/Public+List+of+LAT-Detected+Gamma-Ray+Pulsars}}
\tablenotetext{b}{DM from \cite{2018ATel11584....1W}.}
\tablenotetext{c}{$\gamma$-ray flux from \cite{Guillemot:2013oka}.}
\tablenotetext{d}{DM from \cite{Clark:2018zsp}.}
\tablenotetext{e}{DM from \cite{Sanpa-arsa:2016}.}
\tablenotetext{f}{DM from 
\url{http://astro.phys.wvu.edu/GalacticMSPs/GalacticMSPs.txt}.}
\end{deluxetable}

\section{Coordinate transformation}
\label{sec:coordinates}
\subsection{Disk profile}
\label{sec:disk_coord}
The number density of MSPs in the disk is parametrized by a cylindrically-symmetric profile. However, observations of MSPs only provide a location on the sky and sometimes distance information. In order to test a particular spatial profile against observations it is useful to convert the number density distribution of MSPs into a probability of finding a source at Galactic coordinates ($\ell, b$) and at a particular distance $D$, $P(\ell, b, D)$ (this function is differential in $\ell$, $b$ and $D$, and normalized to one).

The disk profile is centered on the Galactic center. As a first step, it is useful to convert the cylindrical coordinates into Cartesian coordinates, with the Galactic center at origin:
\begin{equation}
\begin{split}    
    x_\mathrm{GC}(r, z, \theta) &= r\cos\theta \\
    y_\mathrm{GC}(r, z, \theta) &= r\sin\theta \\
    z_\mathrm{GC}(r, z, \theta)  &= z.
\end{split}
\end{equation}
We define the Sun to be at $(x_\mathrm{GC}, y_\mathrm{GC}, z_\mathrm{GC}) = (r_\odot, 0, 0)$. A simple translation suffices to move the sun 
to the origin. We refer to this heliocentric-coordinate system with $(x, y, z)$.
The coordinate system with $(\ell, b, D)$ is related to the Cartesian coordinates through:
\begin{equation}
\begin{split}    
    x_\mathrm{GC}(\ell, b, D) &= r_\odot - D \cos(\ell)\cos(b) \\
    y_\mathrm{GC}(\ell, b, D) &= D \sin(\ell) \cos(b) \\
    z_\mathrm{GC}(b, D) &= D \sin(b).
\end{split}
\end{equation}
With this information we can calculate the relevant Jacobians to perform the coordinate transformation, which for simplicity we perform in two steps to obtain
\begin{equation}
    P(\ell, b, D) = 
    \overbrace{\left| \frac{\partial(x, y, z)}{\partial(\ell, b, D)}\right|}^{=D^2 \cos(b)}
    \overbrace{\left| \frac{\partial(r, z, \theta)}{\partial(x, y, z)}\right|}^{=1/r}
    P(r,z,\theta).
\end{equation}
Note that the Jacobian for the transformation of cylindrical to Cartesian coordinates
cancels the presence of the same term in $P(r,z,\theta)$ described in the main text, Sec.~\ref{sec:rho}.

%
\subsection{Bulge profile}
The coordinate transformation for the bulge profile is analogous to 
that of the disk described in Sect.~\ref{sec:disk_coord}, with the only difference
that we now start from spherical instead of cylindrical coordinates.
In this case,
\begin{equation}
 \left| \frac{\partial(r, \theta, \phi)}{\partial(x, y, z)}\right| = r^{-1}\sin^{-1}\phi.
\end{equation}

\section{Details about model likelihoods}
\subsection{Derivation of the Likelihood Function}
\label{sec:likelihood}
We start with Bayes rule \citep[e.g.,][]{Trotta:2008qt} with $\mathbf{\Theta}$ being our parameters of interest,
\begin{equation}
P(\mathbf{\Theta}|\mathcal{D}) \propto \mathcal{L}(\mathcal{D}|\mathbf{\Theta})\pi(\mathbf{\Theta}),
\end{equation}
where the posterior on the left-hand-side is understood to be normalized to one w.r.t.~$\mathbf{\Theta}$.
Next, we introduce the unbinned likelihood
\begin{equation}
\mathcal{L}(\mathcal{D}|\mathbf{\Theta}) = e^{-\mu(\mathbf{\Theta})}
\prod_{k=1}^3 P(\mathcal{D}_k|\mathbf{\Theta}).
\end{equation}
The product arises because we have three independent datasets \citep[e.g.~][]{Hobson:2002zf}, 
for sources with parallax measurements, sources with dispersion measures, and sources without measurement of a distance proxy. 
Let us next focus on the case where we have a distance measurement,
denoted by $\kappa$,
through either parallax or dispersion measure.
Also, let us make explicit the dependence of the measured values of the distance proxy
and flux on the true distance, flux and their uncertainties ($D,\sigma_\kappa, F$ and
$\sigma_F$). We note that measured spatial
positions are assumed to correspond to the true values in this work.
For a single pulsar, denoted by subscript $i$,
we can use conditional probabilities to write 
\begin{widetext}
\begin{equation}
\begin{split}
P(\ell_i, b_i, F_i, \kappa_i|\mathbf{\Theta}, D, \sigma_\kappa, F, \sigma_F) 
&=
P(\ell_i, b_i|\mathbf{\Theta})P(F_i|F, \sigma_F)  P(\kappa_i|D, \sigma_\kappa) \\
P(\ell_i, b_i, F_i, \kappa_i|\mathbf{\Theta},\sigma_\kappa,\sigma_F) &=\int dF dD P(\ell_i, b_i, D, F|\mathbf{\Theta}) P(F_i|F, \sigma_F) P(\kappa_i|D, \sigma_\kappa)\\
&=
4\pi \int dF dD D^2 P(\ell_i, b_i, D|\mathbf{\Theta}) 
P(L=4\pi D^2F|\mathbf{\Theta}) P(F_i|F, \sigma_F) 
P(\kappa_i|D, \sigma_\kappa).
\end{split}
\end{equation}
\end{widetext}
In the second line we dropped any dependencies on the the spatial position, however,
this can be thought of as being included in the uncertainties ($\sigma_\kappa, \sigma_F$).
In reality the uncertainties
can be a complicated function of the true flux, true distance and spatial position.
Introducing these dependencies is beyond the scope of the current work.
In practice,
we took the uncertainties from the observations, which will implicitly depend on spatial position,
distance and/or flux.
In the second line we also made use of $P(A) = \int P(A, B) dB$ in order to integrate over
distance and flux uncertainties.
Finally, in the third line we changed variables from flux to luminosity. Since luminosity and
spatial position are independent we can write $P(\ell, b, D, L) = P(\ell, b, D)P(L)$.
This reproduces the likelihood in Eq.~\ref{eq:prob} without the flux threshold which is
independent of the above discussion.

\subsection{Luminosity Function Measurement Error}
\label{sec:L_err}
In Fig.~\ref{fig:dNdL_bench} we show the best-fit luminosity function for
our benchmark model and compare it to observations. The data points are expectation
values for each bin 
taking into account the uncertainty in the distance to the sources and on the flux. 
A description on how to compute the expectation values is given below.  We emphasize that this approach is \emph{not} used for the purpose of statistical inference, but \emph{only} for the purpose of facilitating a visual comparison between predicted and measured luminosity functions in Fig.~\ref{fig:dNdL_bench}.

Let us again denote the true luminosity and flux of, and distance to a particular
pulsar by $L, F$ and $D$ and measurements with subscript $i$. In this case we
have for pulsars with a distance proxy ($\kappa_i$):
\begin{widetext}
\begin{equation}
\begin{split}
  P(L|F_i, \kappa_i, \ell, b,\mathbf{\Theta}) &= 
  \int dF dD P(L, D, F|F_i, \kappa_i, \ell, b,\mathbf{\Theta})  \\
  &= \int dF dD P(L|D, F) P(F|F_i, \ell, b) P(D|\kappa_i, \ell, b,\mathbf{\Theta})\;.
\end{split}
\end{equation}
\end{widetext}
Here $P(L|D, F) = \delta\left(L - 4\pi D^2 F\right)$. $P(F|F_i)$ is a Gaussian
similar to that described in Sect.~\ref{sec:flux}, but with the true and 
observed distance interchanged. The probability of a true distance given
the observation is given by Eq.~6 in \cite{Verbiest:2012kh} in case
of a parallax measurement. When only a dispersion measure is available
we derive $P(D|\kappa_i)$ using a Monte-Carlo similar to the one described 
in Sect.~\ref{sec:dist_dm}, but this time obtaining the distance
by sampling from $10^4$ random realizations
of the YMW16 model with the dispersion measure set equal to the measured value.
Finally, when neither parallax nor distance information is present
we use $P(D|\ell, b,\mathbf{\Theta})$ as given in Sect.~\ref{sec:rho}.
We then obtain the expectation value in a particular bin ($L_{-}\leq L < L_+$) through
\begin{widetext}
\begin{equation}
  P\left(L_{-}\leq L < L_{+}|F_i, \kappa_i, \ell, b\right) = 
  \int_{L_{-}\leq 4\pi D^2 F}^{4\pi D^2 F < L_{+}} dF dD P(F|F_i, \ell, b) P(D|\kappa_i,\ell, b).
\end{equation}
\end{widetext}
Contributions from all pulsars are summed to obtain the overall expectation. 
Although the expectations in general are not integers,
errors are treated as Poissonian and so the errorbars
correspond to the square-root of the expectation.

\section{Results for different models}
\label{sec:other_results}
\begin{figure*}[t]
  \centering
  \includegraphics[width=0.99\linewidth]{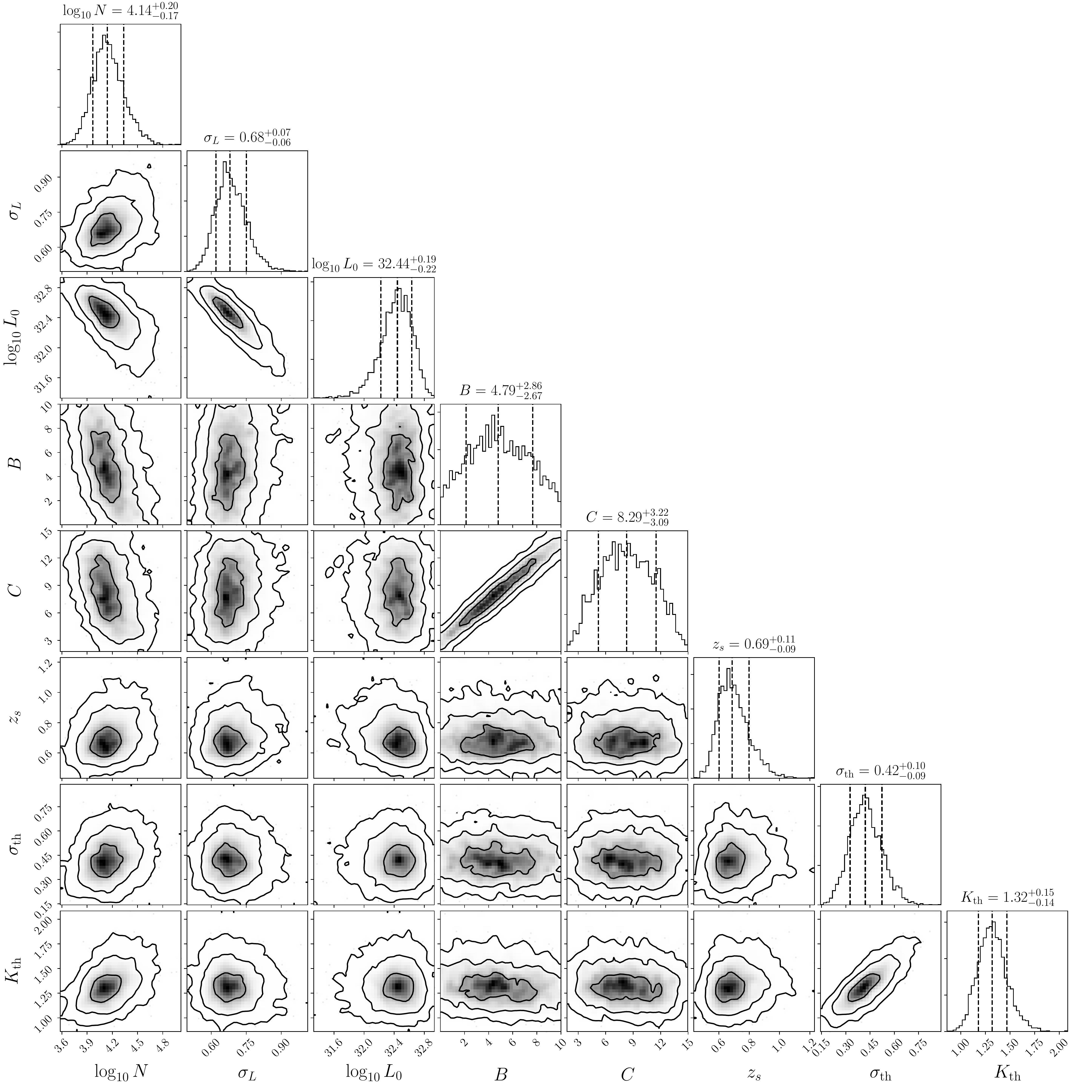}
  \caption{Similar to Fig.~\ref{fig:bench}, but for the log-normal
  luminosity function.}
  \label{fig:cp_ln}
\end{figure*}

\begin{figure*}[t]
  \centering
  \includegraphics[width=0.99\linewidth]{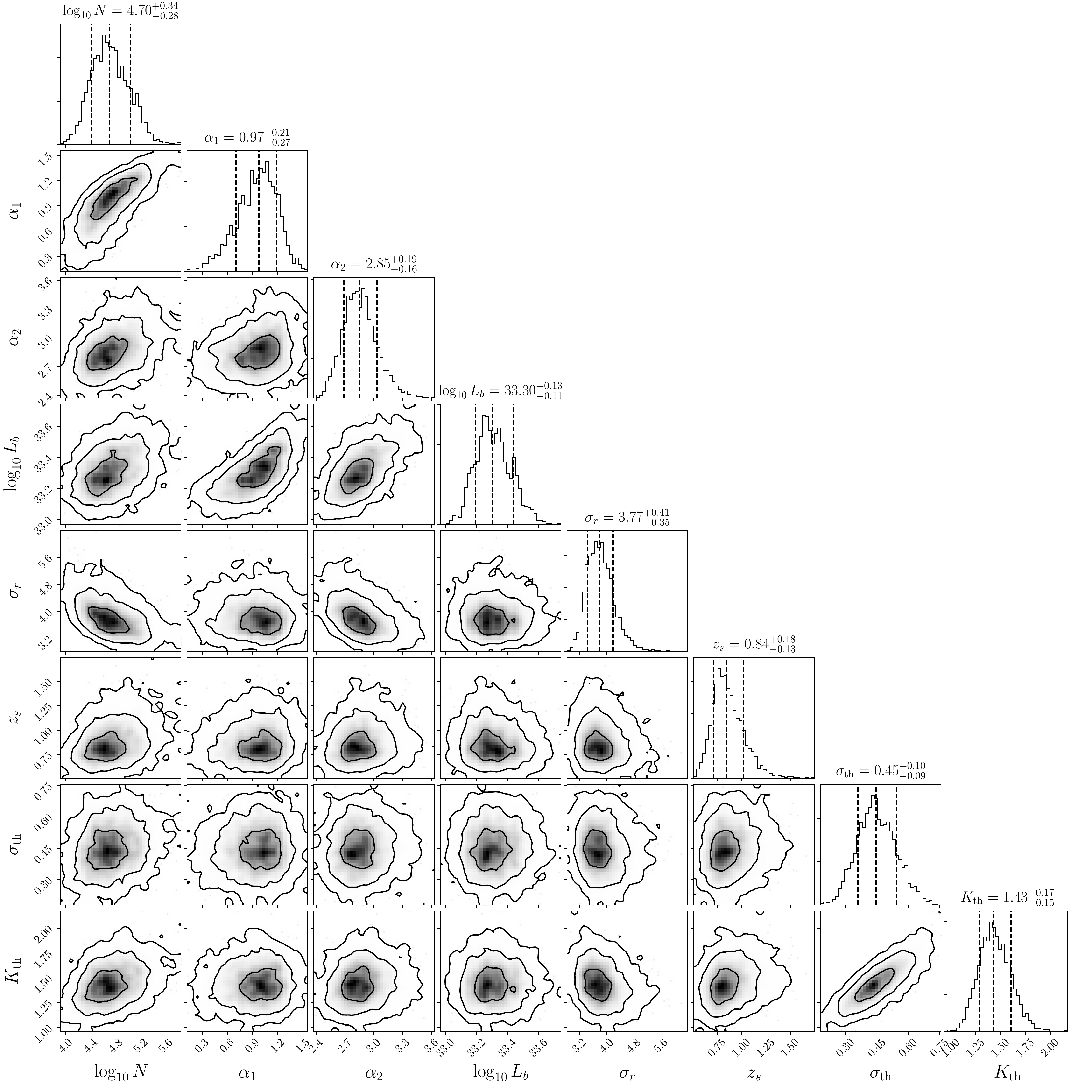}
  \caption{Similar to Fig.~\ref{fig:bench}, but using a 
  Gaussian radial profile for the disk}
  \label{fig:cp_gauss}
\end{figure*}

\begin{figure*}[t]
  \centering
  \includegraphics[width=0.99\linewidth]{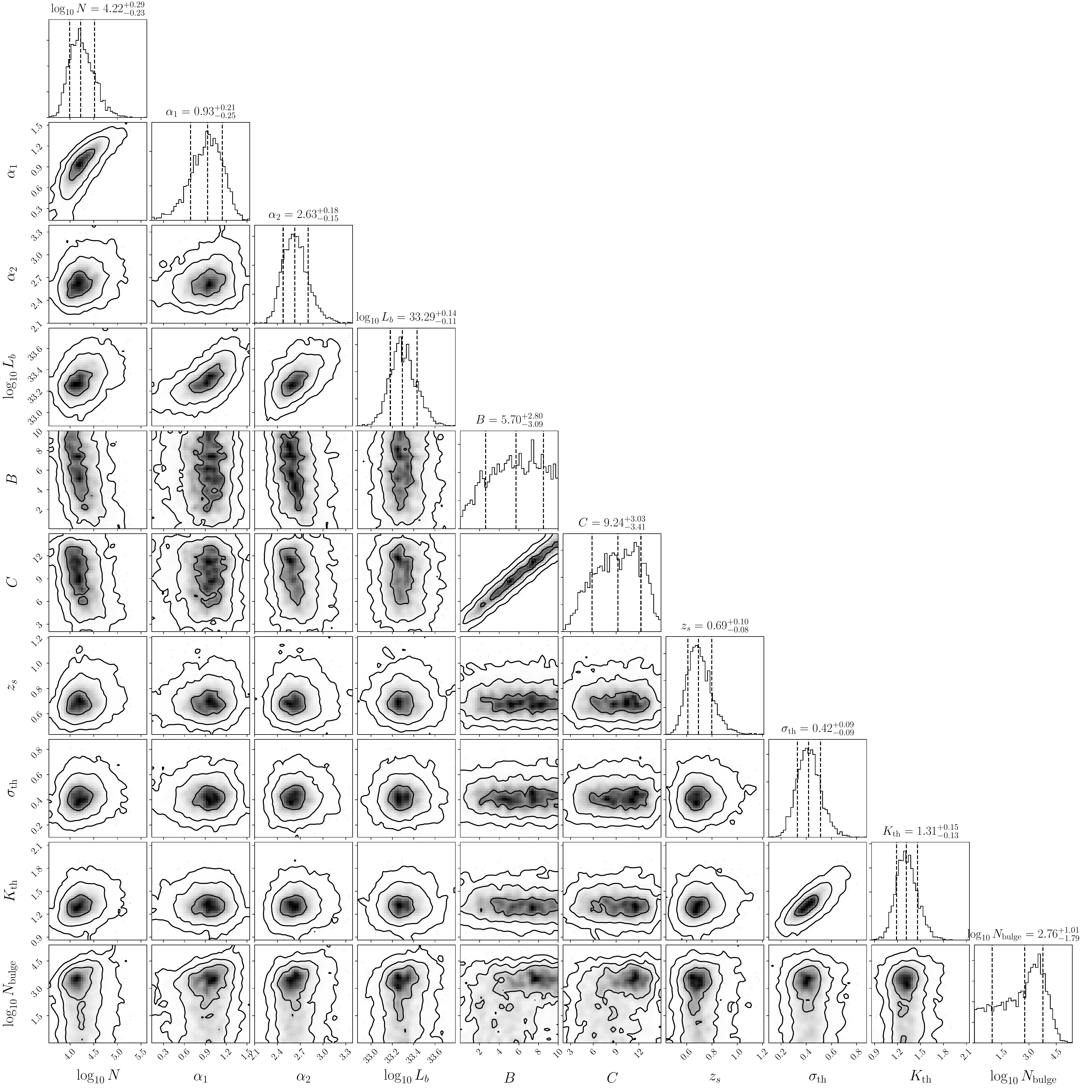}
  \caption{Similar to Fig.~\ref{fig:bench}, but including
  an additional bulge component in the center with an identical luminosity function.}
  \label{fig:cp_bulge}
\end{figure*}

\begin{figure*}[t]
  \centering
  \includegraphics[width=0.99\linewidth]{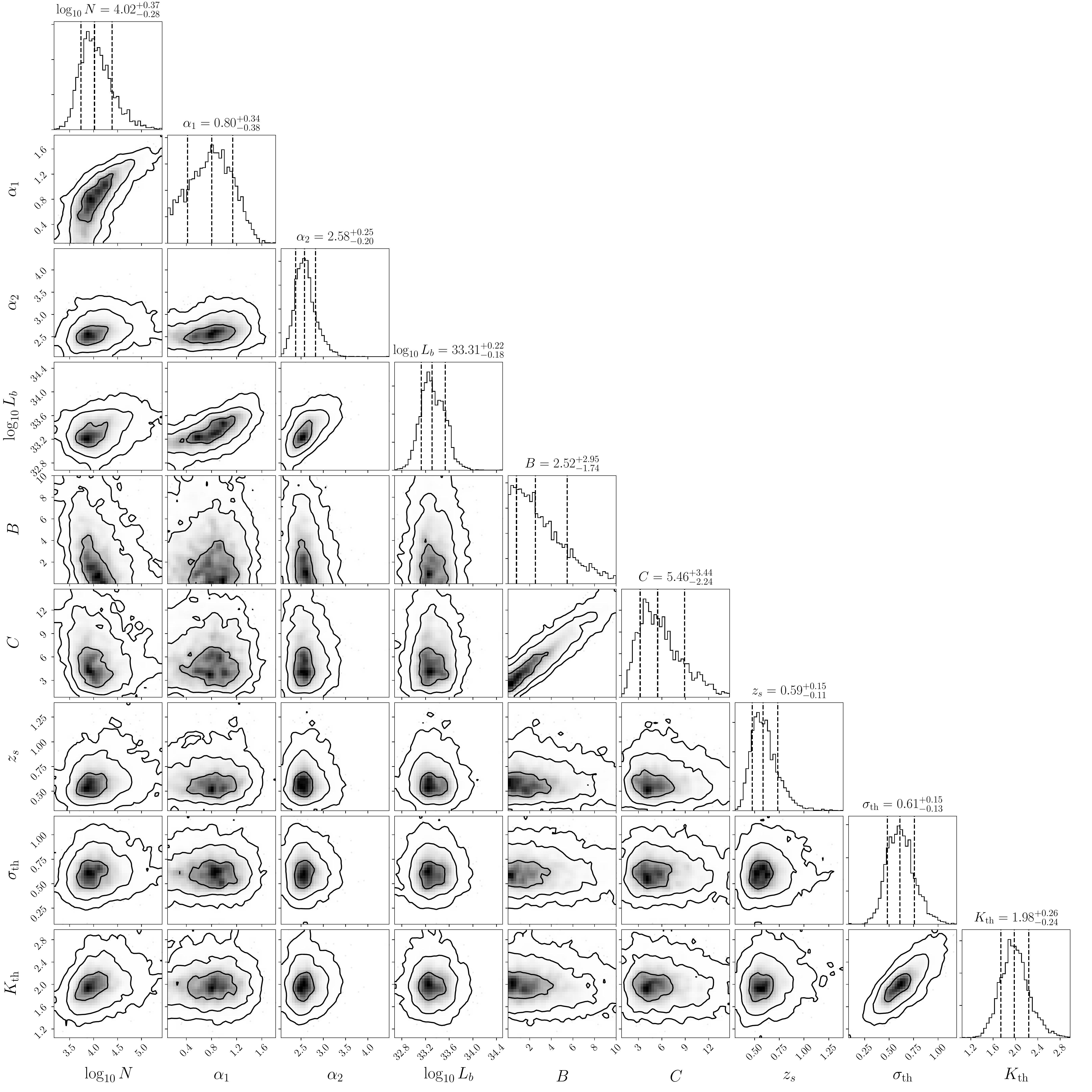}
  \caption{Similar to Fig.~\ref{fig:bench}, but using
  a smaller dataset based on the MSPs in the 2PC \citep{TheFermi-LAT:2013ssa}.}
  \label{fig:cp_2PC}
\end{figure*}

\begin{figure*}[t]
  \centering
  \includegraphics[width=0.99\linewidth]{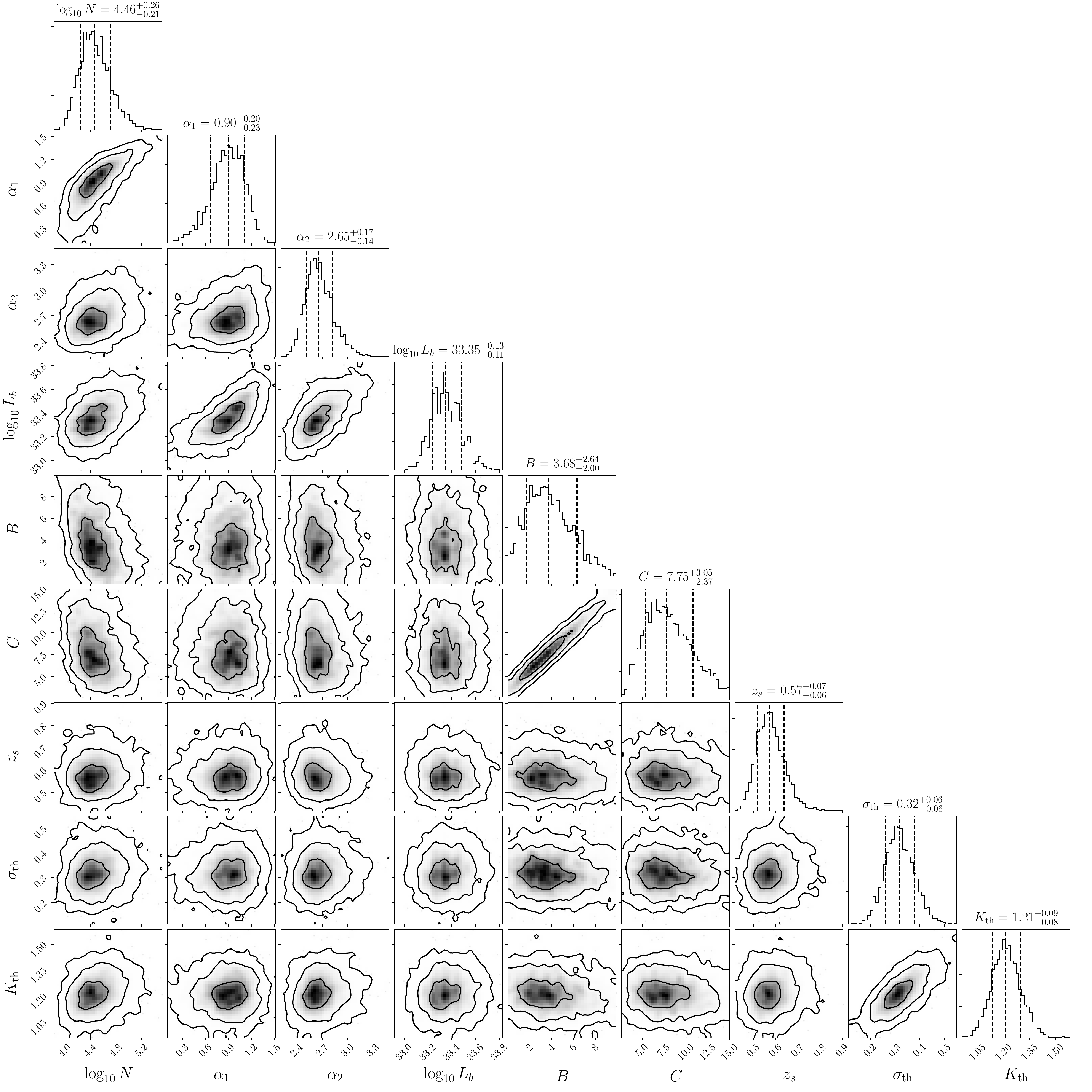}
  \caption{Similar to Fig.~\ref{fig:bench}, but using a larger
  source sample including unassociated sources selected from \cite{Parkinson:2016oab}.}
  \label{fig:cp_ALL}
\end{figure*}

In this section we show corner plots similar to Fig.~\ref{fig:bench} for a selection of 
different models considered in the main text. Only changes 
with respect to the benchmark model are mentioned.
In Fig.~\ref{fig:cp_ln} we show the result for the model with a log-normal luminosity
function. 
Figure~\ref{fig:cp_gauss} contains the results for a model
with a Gaussian disk profile.
The results for a model including a bulge component is
displayed in Fig.~\ref{fig:cp_bulge}.
Finally, we show
the results obtained when using a pulsar sample consisting only of the 2PC
MSPs (Fig.~\ref{fig:cp_2PC}) and with the addition of
unassociated sources (Fig.~\ref{fig:cp_ALL}).

\end{document}